\documentclass[a4paper,11pt]{article}
\pdfoutput=1 

\usepackage{jheppub} 

\usepackage{graphicx}
\usepackage[figurename=Fig.]{caption}
\usepackage[tablename=Tab.]{caption}
\usepackage{color}
\usepackage{float}
\usepackage{amssymb}
\usepackage{amsmath}
\usepackage{textcomp}
\usepackage{cases}
\usepackage{makecell}
\usepackage{pdflscape}
\usepackage{subfigure}
\usepackage{pdfpages}
\usepackage{makeidx}
\usepackage{bm}
\usepackage{xstring}
\usepackage{lipsum}
\usepackage{slashed}
\usepackage{multicol}
\usepackage{setspace}
\usepackage{booktabs}
\usepackage[table]{xcolor}
\usepackage{multirow}
\usepackage{braket}
\usepackage{esvect}[e]
\usepackage[separate-uncertainty]{siunitx}
\usepackage[displaymath, mathlines]{lineno}

\newcommand{\eref}[1]{Eq.\,(\ref{#1})}
\newcommand{\w}{\omega}
\newcommand{\vo}{\vec{o}\@ifnextchar{^}{\,}{}}
\def\up{\mathrm}
\def\d{\mathrm{d}}

\def\i{\mathrm{i}}

\renewcommand{\thetable}{\Roman{table}}
\newcommand{\vabs}[1]{|\vec #1\,|}

\newcommand{\overlrarrow}[1]{{\mathop{#1} \limits ^{\leftrightarrow}}}
\newcommand{\comment}[1]{}

\graphicspath{{PDF/}}

\def\slash#1{\setbox0=\hbox{$#1$}           
   \dimen0=\wd0                                 
   \setbox1=\hbox{/} \dimen1=\wd1               
   \ifdim\dimen0>\dimen1                        
      \rlap{\hbox to \dimen0{\hfil/\hfil}}      
      #1                                        
   \else                                        
      \rlap{\hbox to \dimen1{\hfil$#1$\hfil}}   
      /                                         
   \fi}                                         %
\def\sl#1{\setbox0=\hbox{#1}
  \dimen0=\wd0
  \rlap{\hbox to \dimen0{\hss/\hss}}%
  #1}

\title{\boldmath Strong decays of $P_{\psi}^N(4312)^+$ to $J/\psi(\eta_c) p$ and $\bar D^{(*)}\Lambda_c$ within the Bethe-Salpeter framework}

\author[1]{Qiang Li}
\author[2,3,4]{Chao-Hsi Chang}
\author[5]{Tianhong Wang}
\author[5,6]{Guo-Li Wang}

\affiliation[1]{School of Physical Science and Technology, Northwestern Polytechnical University, Xi'an 710072, China}
\affiliation[2]{CAS Key Laboratory of Theoretical Physics, Institute of Theoretical Physics, Chinese Academy of Sciences, Beijing 100190, China}
\affiliation[3]{School of Physical Sciences, University of Chinese Academy of Sciences, Beijing 100049, China}
\affiliation[4]{School of Physical Science and Technology, Lanzhou University, Lanzhou 730000, China}
\affiliation[5]{School of Physics, Harbin Institute of Technology, Harbin 150001, China}
\affiliation[6]{Department of Physics, Hebei University, Baoding 071002, China}

\emailAdd{liruo@nwpu.edu.cn}
\emailAdd{zhangzx@itp.ac.cn}
\emailAdd{thwang@hit.edu.cn}
\emailAdd{gl\_wang@hit.edu.cn}

\abstract{
Based on the effective Lagrangian in the heavy quark limit, we calculate the one-boson-exchange interaction kernel of $P_\psi^N(4312)^+$ as the $\bar D\Sigma_c$ molecular state in isospin-$\frac12$. We present the Bethe-Salpeter equation and wave function for the constituent particles to be a (pseudo)scalar meson and a $\frac12$ baryon.  By solving the Bethe-Salpeter equation, we obtain  $P_\psi^N(4312)^+$ as the $\bar D\Sigma_c$ molecular state  with $J^P=(\frac{1}{2})^-$. Combining the effective Lagrangian and the obtained BS wave function, the partial decay widths of  $P_\psi^N(4312)^+$ to $J/\psi p$,  $\eta_c p$, $\bar D^{*0}\Lambda_c^+$ and $\bar D^0\Lambda_c^+$ are calculated to be $0.17$, $0.085$, $8.8$, and $0.026$ MeV, respectively, which are roughly consistent with the LHCb experimental measurements and some other theoretical researches. The obtain results indicate the fraction of $\bar D^{*0}\Lambda_c^+$ channel amounts to $\sim90\%$ of $P_\psi^N(4312)^+$, and is a highly promising channel to be discovered in the near future experiments. Our results favor the interpretation of  $P_\psi^N(4312)^+$  as the $\bar D\Sigma_c$ molecular state with $J^P=(\frac{1}{2})^-$  and isospin $I=\frac12$.

}

\begin{document}
\maketitle
\flushbottom

\section{Introduction}\label{Sec-1}

In 2019,  a narrow pentaquark state $P_c(4312)^+$ is first observed in the $J/\psi p$ invariant mass spectrum\,\cite{LHCb2019-Pc} by the LHCb collaboration, which indicates this state at least to contain five valence quarks, namely, $[c\bar c uud]$ quark contents. This pentaquark state will be labeled as $P_{\psi}^N(4312)^+$ in this work following the new naming scheme proposed by the LHCb collaboration\,\cite{LHCb2022-Naming}, where the superscript $N$ denotes the isotopic spin $I=\frac12$ and the subscript $\psi$ denotes the hidden charm flavor. The measured mass and total width are $M_{P_{\psi}^N(4312)^+}=4312$ MeV and $\Gamma_{P_\psi^N(4312)^+}=9.8\pm2.7^{+3.7}_{-4.5}$ MeV\,\cite{LHCb2019-Pc} respectively. The proximity to the $\bar D\Sigma_c$ threshold of the observed narrow peak suggests that they play an important role in the dynamics of  ${P_\psi^N(4312)^+}$ state, and makes the  $\bar D\Sigma_c$ molecular state picture a natural interpretation to this exotic particle.

The hidden charm molecular pentaquark states have been proposed before the experimental confirmation\,\cite{WuJJ2010,WangWL2011,WuJJ2012,YangZC2012,LiXQ2014,Karliner2015,ChenR2015}.
After the LHCb discoveries, lots of literature explored these newly observed pentaquark states from different aspects within different approaches, such as Refs.\,\cite{XiaoCJ2019,LiuMZ2019,ChenR2019,XiaoCW2019,HeJ2019,LinYH2019,ChenHX2019,Ali2019,MengL2019,Burns2019,Voloshin2019,GuoFK2020,KeHW2020,DuML2020,WangZG2020,Yamaguchi2020,XuH2020,Burns2022}. Although the properties of the $P_\psi^N(4312)^+$ are most likely to be the $S$-wave combination of $\bar D\Sigma_c$ with $I(J^P)=1/2(1/2^-)$\,\cite{ChenR2019,ChenHX2019,LiuMZ2019,XiaoCJ2019,XiaoCW2019,HeJ2019,LinYH2019,XuH2020,Burns2022}, the contrary view\,\cite{Fernandez-Ramirez2019}, or the possibilities of the compact pentaquark state\,\cite{Ruangyoo2021,Stancu2021} or kinematical effects\,\cite{Nakamura202103,Nakamura202109} still exist.  Though suggested by the LHCb to be labeled as $P_\psi^N(4312)^+$, the essence of this pentaquark state is still an open question.

Besides the spectrum  or electromagnetic properties\,\cite{Ozdem2021A,Ozdem2021,XuYJ2021}, the strong decay properties play important roles in determining the nature of the pentaquark states. The decay to $J/\psi p$ is the discovery channel and also the only detected decay mode of $P_\psi^N(4312)^+$ so far, and hence this decay channel should be paid more attention to explore the property of $P_\psi^N(4312)^+$. Several approaches are used to study the decay properties of these pentaquark states\,\cite{XiaoCJ2019,LinYH2019,Stancu2021,Sakai2019,DongYB2020,WangGJ2020,ChenHX2020,XuYJ2020,WangZG2020A}, including the effective Lagrangian methods\,\cite{XiaoCJ2019,LinYH2019}, the flavor-spin and heavy quark spin symmetry\,\cite{Stancu2021,Sakai2019}, the chiral
constituent quark model\,\cite{DongYB2020}, QCD sum rules\,\cite{WangZG2020A,XuYJ2020}, etc. Most of the previous studies are based on the nonrelativistic Schrodinger or Lippmann-Schwenger equation and the results are dependent on several introduced free parameters, especially the cutoff value in the form factors. These undetermined parameters weaken the prediction power of the theories and bring ambiguity in interpreting the nature of $P_\psi^N(4312)^+$. Some researches also suggest the $\eta_c p$ channel can be an important decay mode of $P_\psi^N(4312)^+$\,\cite{DongYB2020,WangGJ2020}, especially, the methods by using the heavy quark symmetry predict that the decay ratio of $P_\psi^N(4312)^+$ to $\eta_c p$ over $J/\psi p$ can reach about three\,\cite{Voloshin2019,Sakai2019}. However,  no experimental evidence is reported in a recent search for pentaquark state $P_\psi^N(4312)^+$  in $\Lambda_b^0 \to \eta_c p K^-$ decay channel\,\cite{LHCb2020-Lambdab}. Also some studies suggest the $\bar D^{*0}\Lambda_c^+$ may be the dominant decay channels of $P_\psi^N(4312)^+$\,\cite{LinYH2019,DongYB2020}. Though there are already some calculations on these decays, the relevant researches are still relatively scarce and the current predictions are not well consistent with each other.  More studies on the decay behaviors of $P_\psi^N(4312)^+$ can be important and helpful to explore its inner structure and dynamics.

In this work, we will calculate the partial decay widths of $P^N_\psi(4312)^+$ to $J/\psi (\eta_c) p$ and $\bar D^{(*)0}\Lambda_c^+$ by combing the Bethe-Salpeter\,(BS) framework  with the effective Lagrangian.  The Bethe-Salpeter equation(BSE) is a relativistic two-body bound state equation. Another advantage is that the constructed BS wave functions only depend on the good quantum number spin-parity and Lorentz covariance. The BS methods have already been successfully used to cope with mass spectra of the doubly heavy baryons\,\cite{LiQ2020,LiQ2022},  producing the recently observed molecular  pentaquarks\,\cite{XuH2020} and the fully heavy tetraquark $ T_{QQ\bar Q\bar Q}$ states\,\cite{LiQ2021}, and also the hadronic transitions and decays\cite{Chang2005,WangZ2012A,WangT2013,LiQ2016,LiQ2017,LiQ2017A,LiQ2019A}. The theoretical calculations from BS methods achieve satisfactory consistences with the experimental measurements. 

This paper is organized as follows. After the introduction, we start with the Bethe-Salpeter equation for $P_\psi^N(4312)^+$ as the molecular state of a (pseudo)scalar meson and a baryon, including  the interaction kernel and the relevant Salpeter wave function (Sect. \ref{Sec-2}), then we calculate the strong decay widths of $P_\psi^N(4312)^+\to J/\psi(\eta_c) p$ and $\bar D^{(*)0}\Lambda_c^+$ (Sect.\,\ref{Sec-3}). We finally present the numerical results, discussion and summaries in Sect.\,\ref{Sec-4}.

\section{$P_{\psi}^N(4312)^+$ as the $\bar D \Sigma_c$ molecular state}\label{Sec-2}

In this part, we will first briefly review Bethe-Salpeter equation of  a scalar meson and a baryon under the instantaneous approximation. Then we introduce the pentaquark interaction kernel based on the one-boson exchange. The relativistic BS wave functions of the $J^P=(\frac12)^-$ $P_\psi^N$ state will be introduced and solved numerically to prepare for  the next decay calculations.
\begin{figure}[htpb]
\centering
\includegraphics[width = 0.7\textwidth, angle=0]{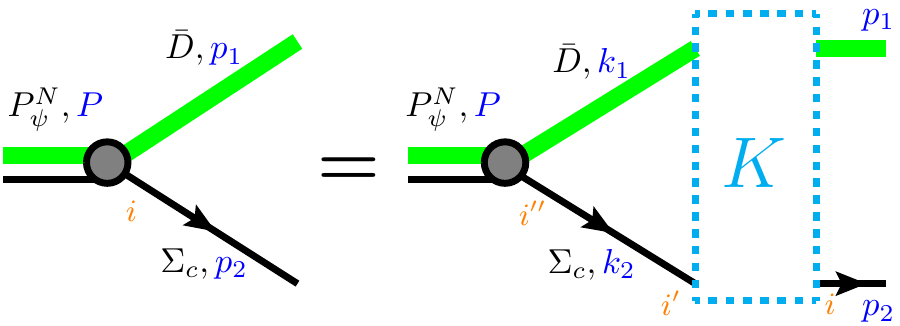}
\caption{Bethe-Salpeter equation of the molecular state consisting of the constituent (pseudo)scalar meson and $J=\frac12$ baryon.  The orange letters denote the Dirac indices. The blue symbols $P,~p_1(k_1),~p_2(k_2)$ denote the momenta of the pentaquark, constituent meson, and the constituent baryon respectively.}\label{Fig-BS-Pc}
\end{figure}

\subsection{Bethe-Salpeter equation of  a $J^P=0^-$ meson and a $\frac{1}{2}^+$ baryon}

\autoref{Fig-BS-Pc} schematically depicts the Bethe-Salpeter equation for the bound state consisting of a constituent meson and a constituent baryon, which can be expressed as
\begin{gather} \label{E-BSE-Pc}
\Gamma(P,q,r) =\int \frac{\up d^4 k}{(2\pi)^4} (-i)K(k,q)  [S(k_2) \Gamma(P,k,r) D(k_1)] ,
\end{gather}
where $\Gamma(P,q,r)$ denotes the vertex of the pentaquark, constituent meson and baryon; we used $P$, $q$, and $r$ to represent the pentaquark total momentum, inner relative momentum, and spin state respectively; the inner relative momentum $q$  and $k$ are defined as $q\equiv\alpha_2p_1-\alpha_1p_2$, $k\equiv\alpha_2k_1-\alpha_1k_2$, with $\alpha_{1(2)}\equiv \frac{m_{1(2)}}{m_1+m_2}$, $k_{1(2)}$ denoting the momentum of the constituent meson\,(baryon), and $m_{1(2)}$ is the corresponding mass; $S(k_2)=\frac{\i}{\slashed k_{\!2} - m_2}$ is the free Dirac propagator of the baryon; $D(k_1)=\frac{\i}{k_1^2-m_1^2}$ denotes the usual scalar propagator. The $\i \epsilon$ should be implied in all the propagators. Since both the two constituent particles, namely $\bar D$ and $\Sigma_c$, contain a heavy charm quark, the relative velocity would be small.  Then the interaction kernel $K(k,q)$ is assumed to be instantaneous and is not dependent on the time component of the exchanged momentum $(k-q)$, namely, $K(k,q)\sim K(k_\perp-q_\perp)$, where $k_\perp = k - k_P \hat P$ with $k_P\equiv k\cdot \hat P$ and $\hat P = \frac{P}{M}$, and $M$ is the pentaquark mass. The spacelike momentum $q_\perp$ is defined similarly. Throughout this work, this instantaneous approximation is assumed for the pentaquark kernel.

The four-dimensional Bethe-Salpeter wave function is  defined as
\begin{gather} \label{E-wave-4D}
\psi (q)=S(p_2) \Gamma(q) D(p_1),
\end{gather}
where the dependence on $P$ and $r$ omitted for simplicity.
Since the interaction kernel $K(k_\perp-q_\perp)$ is instantaneous, the integral over the time component of $q$ can be absorbed into the wave function and  it is useful to define the three-dimensional Salpeter wave function as
\begin{gather}
 \varphi(q_\perp) \equiv - \i \int \frac{\up d q_P}{2\pi} \psi(q).
\end{gather}
where the factor $(-\i)$ is just a convention for later convenience.

Performing the contour integral over $q_P$ on both sides of \eref{E-wave-4D}, we can obtain the Salpeter equation (SE) for meson-baryon bound state\,\cite{XuH2020},
\begin{align} \label{E-SE-0}
\varphi(q_\perp)=\frac{1}{2w_1}\left[ \frac{\Lambda^+(p_{2\perp}) }{M-w_1-w_2} +  \frac{\Lambda^-(p_{2\bot})  }{M+w_1+w_2}  \right] \Gamma(q_\perp),
\end{align}
where $w_{i}=\left({m_{i}^2-p_{i\perp}^2}\right)^{1/2}~(i=1, 2)$ denotes the kinetic energy of the constituent meson and baryon respectively.  The projector operators $\Lambda^\pm(p_{2\perp})$ are defined as 
\begin{gather}
\Lambda^\pm(p_{2\perp})=\frac{1}{2}\left[1 \pm {H}_2(p_{2\perp})\right]\gamma_0,\\
H_2(p_{2\perp})=\frac{1}{w_2} \left(\slash p_{2\perp}+m_2 \right)\gamma^0.
\end{gather}
Notice that $H_2(p_{2\perp})$ is just the corresponding Dirac Hamiltonian divided by the kinetic energy $w_2$.
Using the projector operator, we can further define the positive and negative energy wave functions as $\varphi^{\pm} \equiv \Lambda^\pm  \gamma^0 \varphi $, and we also have $\varphi=\varphi^+ + \varphi^-$. The SE above can be further rewritten as the following  type
\begin{gather} \label{E-VS-SE}
M \varphi =  (w_1+w_2)  H_2(p_{2\perp})\varphi +\frac{1}{2w_1} \gamma_0 \Gamma(q_\perp).
\end{gather}
where the vertex $\Gamma(q_\perp)$ is now expressed as the integral of the Salpeter wave function, 
\begin{align} \label{E-Gamma-3D}
\Gamma(q_\perp) =\int \frac{\up d^3 k_\perp}{(2\pi)^3}  K(k_\perp-q_\perp)  \varphi(k_\perp) .
\end{align}
The  Salpeter  \eref{E-VS-SE} is in fact an eigenvalue equation about the Salpeter wave function $\varphi_\alpha(q_\perp)$, where the pentaquark mass $M$ behaves as the eigenvalue. The three-dimensional BSE, namely, \eref{E-VS-SE}, indicates that the mass of the pentaquark state consists of two parts, the kinetic energy and the potential energy. 

The normalization condition of the BS wave function is generally expressed as,
\begin{align*}
&-\i \int \int \frac{\up{d}^4 q}{(2\pi)^4}\frac{\up{d}^4 k}{(2\pi)^4} \bar{\psi}(P,q, \bar r) \frac{\partial}{\partial P^0} I(P,k,q) \psi(P,k,r)  = 2M \delta_{r\bar r},
\end{align*}
where $\bar \psi = \psi^\dagger\gamma^0$ and $\delta_{r\bar r}$ is the Kronecker symbol; the integral kernel $I$ in the normalization condition reads,
\begin{align*}
I(P,k,q)&=(2\pi)^2 \delta^4(k-q) S^{-1}(p_2)D^{-1}(p_1) + \i K(P,k,q).
\end{align*}
Notice in this work, under the instantaneous approximation, the interaction kernel has no dependence on $P^0$, which indicates the normalization would only involve the inverses of the two propagators. Performing the contour integral, the normalization condition can be further expressed by the Salpeter wave function as
\begin{align}\label{E-Norm1-D1}
\int \frac{\up{d}^3 q_\perp}{(2\pi)^3} (2w_1)  \bar\varphi(q_\perp,\bar r) \gamma^0 \varphi(q_\perp,r)=2M\delta_{r\bar r},
\end{align}
where $\bar \varphi=\varphi^\dagger\gamma^0$ and the symbol $\bar r$ just denotes the spin state; also notice both the BS wave function $\psi(q)$ and the Salpeter wave function $\varphi(q_\perp)$ are four-component spinor.

\subsection{Interaction kernel from the one-boson exchange}

The $P_{\psi}^N(4312)^+$ are consistent with the  $\bar D \Sigma_c$ molecular state with  isospin $I=\frac12$ and $I_3=+\frac12$, which can be expressed in the uncoupled representation as
\begin{gather}
\textstyle \ket{\frac12, \frac12} =\frac{ \sqrt{2}}{ \sqrt{3}} \ket{1,+1}\ket{\frac12,-\frac12} - \frac{1}{\sqrt3} \ket{1,0}\ket{\frac12,\frac12}=\frac{ \sqrt{2}}{ \sqrt{3}} \ket{\Sigma_c^{++}}\ket{D^-} - \frac{1}{\sqrt3} \ket{\Sigma_c^+}\ket{\bar D^0}.
\end{gather}
In the molecular state scenario of $P_\psi^N(4312)^+$, the interaction kernel between the two constituents  $\Sigma_c$ and $\bar D$ can be realized by the one-boson exchange. Notice the usual one-pion exchange is not possible in the $\bar D\Sigma_c$ bound state for the parity. We only need to consider the light scalar and vector meson exchange. 

Considering the heavy quark spin-flavor symmetry, hidden local symmetry and the light quark chiral symmetry, the involved Lagrangian describing the charmed anti-heavy-light meson and a light scalar and vector meson reads\,\cite{Casalbuoni1997,YangZC2012}
\begin{align}
\mathcal{L}_{H\!H {V}} 
&=  -\rho_\up{V1} \braket{\bar H_{\bar c} v_\alpha  {V}^\alpha H_{\bar c}}- \rho_\up{T1} \braket{ \bar H_{\bar c} \sigma^{\alpha\beta} (\partial_\alpha  {V}_\beta - \partial_\beta  {V}_\alpha) H_{\bar c}}+  \sigma_1 \braket{\bar H_{\bar c} \sigma H_{\bar c}}.
\end{align}
Here $\braket{~}$ denotes taking the Dirac trace, and $\sigma$ denotes field of the light scalar meson. $\rho_{V1}$, $\rho_{T1}$, and $\sigma_1$ denote the corresponding coupling constants. $H_{\bar c}$ represents the  field of the $(\bar D, \bar D^*)$ doublet in the heavy quark limit, 
\begin{gather} \label{E-HQbar}
H_{\bar c} = \left(\bar D^{*\mu}\gamma_\mu +\i \bar D \gamma_5\right) \frac{1-\slashed v}{2},
\end{gather}
where $\bar D=(\bar D^0, D^-, D_s^-)$ denotes anti-charmed heavy-light meson fields in flavor triplet, and $\bar D^{*\mu}$ is the corresponding vector state; $\bar H_{\bar c}=\gamma^0H^\dagger_{\bar c} \gamma_0$ is the usual conjunction in Dirac space; and $v$ denotes four-velocity of the heavy-light meson. The symbol $\mathtt{V}$ denotes the $3\times3$ matrix consisting of the 9 light vector meson fields\,\cite{Casalbuoni1997,YangZC2012}
\begin{gather}
{V}=
\begin{bmatrix}
 \frac{(\rho^0+ \omega) }{\sqrt2} & \rho^+  & K^{*+ }\\
\rho^-  & - \frac{(\rho^0 - \omega) }{\sqrt{2}}   & K^{*0} \\
K^{*-}  & \bar K^{*0}  & \phi
\end{bmatrix}.
\end{gather}

Considering the heavy quark symmetry, hidden local symmetry and chiral symmetry, the effective Lagrangian of the heavy-light baryon and light mesons reads\,\cite{Yan1992,ChengHY2007,LiuYR2012,YangZC2012}
\begin{align}
\mathcal{L}_{ {B_6 B_6 V}} 
&= \rho_\up{V2} \braket{\bar{ {S}}_\mu v_\alpha   {V}^\alpha  {S}^\mu  } + \i \rho_\up{T2}   \braket{\bar{ {S}}_\mu (\partial_\mu  {V}_\nu - \partial_\nu  {V}_\mu)  {S}_\nu} +\sigma_2 \braket{\bar{ {S}}_\mu \sigma  {S}^\mu}.
\end{align}
Here $\langle~\rangle$ denotes taking trace in the $3\times3$ flavor space. The baryon spin doublet are incorporated in field
\begin{gather}
 {S}_\mu = -\frac{1}{\sqrt3} (\gamma_\mu + v_\mu) \gamma^5  {B}_6 +  {B}_{6\mu}^{*},
\end{gather}
where the systematic baryon sextet $ {B}_6$ in $3\times3$ matrix reads
\begin{gather}
 {B}_6=
\begin{bmatrix}
\Sigma_c^{++} & \frac{1}{\sqrt2} \Sigma_c^+  & \frac{1}{\sqrt2} \Xi_c^{\prime+} \\
\frac{1}{\sqrt2} \Sigma_c^+ & \Sigma_c^{0}  & \frac{1}{\sqrt2} \Xi_c^{\prime0} \\
\frac{1}{\sqrt2} \Xi_c^{\prime+} & \frac{1}{\sqrt2} \Xi_c^{\prime0} &\Omega_c^0
\end{bmatrix}.
\end{gather}
The conjugation defines as usual for the spinor field $\bar{ {S}}^{mn}_\mu = ( {S}^{mn}_\mu)^\dagger \gamma_0$. An asterisk on the symbol denotes the corresponding spin-$\frac32$ baryon, which is not involved in this work.

Using above relevant Lagrangian and based on the one-boson exchange, we calculate the interaction kernel of  $\bar D\Sigma_c$ in isospin-$\frac{1}{2}$ as
\begin{gather} \label{E-KPB}
K(s_\perp) = F^2(s_{\!\perp}^2) \left(V_1+ V_2 \frac{\slashed s_{\!\perp}}{|\vec s\,|}\right),
\end{gather}
where $F(s_{\!\perp}^2)$ denotes the regulator in the heavy hadron ($\bar D$ or $\Sigma_c$ here) vertex; and the potential $V_1$ and $V_2$ is specifically expressed as,
\begin{equation}\label{E-KVB-Vn}
\begin{aligned}
V_1&=  - 2 \sigma_1 \sigma_2 M_D \frac{1}{E^2_\sigma} + \rho_\up{V1} \rho_\up{V2}M_D \left(\frac{1}{E^2_\rho} - \frac{1}{2E^2_\w} \right) , \\
V_2&= - \frac{1}{3}\rho_\up{V1} \rho_\up{T2}M_D |\bm s|\left(\frac{2}{E^2_\rho} - \frac{1}{E^2_\w} \right) ,
\end{aligned}
\end{equation}
where $E_\rho=({\bm s^2+m^2_\rho})^{1/2}$ denotes the energy of the inter-mediator $\rho$ meson, and similar for $E_\sigma$ and $E_\w$. The influence of the potential strength on the decay widths will be discussed later.

There is no general method to choose the regulator functions. In this work, we use the following propagator-type form factor, namely,
\begin{gather} \label{E-form-factor}
F(\bm s^2)= \frac{m_\Lambda^2}{\bm{s}^2 +m_\Lambda^2},
\end{gather}
where $m_\Lambda$ is the introduced cutoff parameter to characterize the regulator function. Notice $m_\Lambda$ is the only free parameter in this analysis and can be  determined by fitting bound state mass to the experimental data, which is found to be $m_\Lambda=1.25\,\si{GeV}$ for $P_\psi^N(4312)^+$ and close to the mass scale of the exchanged particle.  In the limit  $\bm{s}^2\to 0$, the heavy hadron is seen by the inter-mediator mesons as a point-like particle, and hence the form factor is normalized to 1.  The cutoff value $m_\Lambda$ is usually believed to be much larger than the typical energy scale  $\sqrt{2\mu\epsilon}\sim0.1\,\si{GeV}$ for $P_\psi^N(4312)^+$\,\cite{GuoFK2018,LinYH2019}, where $\mu=\frac{m_1m_2}{m_1+m_2}$ is the reduced mass of the two-hadron system and $\epsilon=(m_1+m_2-M)$ denotes the bound energy. Our determined cutoff value is consistent with this universal estimation. 
The obtained $V_1$ and $V_2$ for isospin-$\frac{1}{2}$ are displayed graphically in \autoref{Fig-V2}.

\subsection{Salpeter wave function for the $J^P=\frac{1}{2}^-$ pentaquark states}

According to the spin-parity properties, and also considering the proper Lorentz structures,
the Salpeter wave function of $J^P=\frac{1}{2}^-$ pentaquarks consisting of a $0^-$ meson and $\frac{1}{2}^+$ baryon can be generally constructed as
\begin{gather} \label{E-wave1-1-2N}
\varphi(P,q_\perp,r)  =\left (f_1+f_2 \frac{\sl q_\perp}{q} \right) \gamma^5 u(P,r) ,
\end{gather}
where the radial wave function $f_{1(2)}(\vabs{q})$ only explicitly depend on $\vabs{q}$; $u(P,r)$ denotes Dirac spinor with spin state $r$.  In terms of the spherical harmonics $Y_l^m$, the wave function can be rewritten as
\begin{gather} \label{E-wave0-Ylm}
\varphi(P,q_\perp,r)= 2 \sqrt{\pi}\left[f_1 Y_0^0  + \frac{1}{\sqrt{3}}  f_2 {\left( Y_1^{1}\gamma^- + Y_1^{-1} \gamma^+ -Y_1^0 \gamma^3\right) }  \right]\gamma^5 u(P,r),
\end{gather}
where $\gamma^{\pm} = \mp\frac{1}{\sqrt{2}} (\gamma^1\pm \i \gamma^2)$. Then it is obvious to see that $f_1$ and $f_2$ represent the $S$- and $P$-wave components, respectively.
Inserting the wave function into \eref{E-Norm1-D1}, we obtain the normalization satisfied by the radial wave functions as
\begin{align}\label{E-Norm1-D1}
\int \frac{\up{d}^3 q_\perp}{(2\pi)^3} 2w_1 \left(f_1^2 +f_2^2 \right)=1.
\end{align}

Inserting the Salpeter wave function \eref{E-wave1-1-2N} into the Salpeter equation (\ref{E-VS-SE}), eliminating the spinor, calculating the trace, we can obtain two coupled eigenvalue equations with the pentaquark mass $M$ as the eigenvalue and $f_{1(2)}$ as the eigen wave functions (see Ref.\,\cite{LiQ2020,LiQ2022,XuH2020} for details). Solving the eigenvalue equations numerically,  we can obtain the corresponding mass spectra and numerical wave functions, which are also graphically displayed in \autoref{Fig-wave-0}.

\section{Strong decays of $P_\psi^N(4312)^+$ to $J/\psi(\eta_c) p$ and $\bar D^{(*)0}\Lambda_c^+$ within the BS wave function}\label{Sec-3}
In this section, we first present the relevant effective Lagrangian; then we give the decay amplitude by using the BS wave function combining with the effective Lagrangian; finally, the expressions of the partial decay widths are presented in terms of the relevant form factors. 

For $P_\psi^N(4312)^+ \to J/\psi(\eta_c) p$, the involved interactions are $J/\psi DD^{(*)}$, $\eta_c DD^*$, and $\Sigma_c ND^{(*)}$, which involve the Lagrangian of the doubly heavy meson and the heavy-light meson. The heavy-light charmed mesons in $S$-wave can be represented by\,\cite{Wise1992,Burdman1992,Casalbuoni1997}
\begin{gather}\label{E-HQ}
H_{c} = \frac{1+\slashed v}{2} (D^{*\mu} \gamma_\mu +\i D \gamma_5),
\end{gather}
where $D^{*\mu}$ and $D$ denote  the corresponding vector and pseudoscalar charmed $D$ mesons respectively. The anti-heavy-light meson doublet $H_{\bar c}$ has been presented in \eref{E-HQbar}. 

For doubly heavy mesons, the heavy quark flavor symmetry does not hold any longer, while the heavy quark spin symmetry still holds. In the ground states, the charmonium forms a doublet consisting of a pseudoscalar $\eta_c$ and a vector state $J/\psi$, which can be represented by\,\cite{Jenkins1992}
\begin{gather}
R =   \frac{1+\slashed v}{2}   (\psi^\mu \gamma_\mu + \i \eta_c \gamma_5)  \frac{1-\slashed v}{2}  ,
\end{gather}
where $\psi^\mu$ and $\eta_c$  denotes the fields of the corresponding mesons.
Here all the hadron fields in above equations contain a factor of $\sqrt{M_H}$ with $M_H$ the corresponding meson mass.

By assuming the invariance under independent rotations of the heavy quark spins, it is possible to write down the effective coupling between the $S$-wave charmonia and the heavy-light mesons as\,\cite{Colangelo2004}
\begin{gather}
\mathcal{L}_2 = g_2 \up{Tr}\, (  R\bar H_{\bar c}  \overlrarrow{\slashed \partial} \bar H_{c}  ) +\up{H.c.},
\end{gather}
which is invariant under independent heavy quark spin symmetry; and the notation $A\overlrarrow{\partial} B\equiv A\partial B - \partial A B$ is used. Consequently, we obtain the following effective Lagrangian describing $J/\psi$ and $\eta_c$ coupling to the $DD^*$,
\comment{
\begin{equation} \label{E-L-All}
\begin{aligned}
\mathcal{L}_{2} =&+g_{\psi  D D} \psi^{\dagger\mu} \bar D   \overlrarrow{\partial}_\mu D    \\
& - \i g_{\psi DD^*} \frac{1}{M_\psi} \epsilon^{\mu\nu\alpha\beta}  {\partial_\mu \psi^\dagger_\nu}( \bar D \overlrarrow{\partial}_\alpha D^*_\beta   +\bar D^*_\alpha  \overlrarrow{ \partial}_\beta D   ) \\
& +g_{\psi D^* D^*} \psi^{\dagger\mu} (\bar D^{*\nu} \overlrarrow{\partial}_\nu D^*_\mu  + \bar D^*_\mu  \overlrarrow{\partial}_\nu D^{*\nu} - \bar D^{*\nu} \overlrarrow{\partial}_\mu D^{*}_\nu) \\
&+g_{D D^*\eta_c}  \eta^\dagger_c ({\partial_\mu} \bar D D^{*\mu} - \bar D^{*\mu} {\partial_\mu} D) \\
& + \i  g_{D^* D^*\eta_c} \frac{1}{M_{\eta_c}}\epsilon^{\mu\nu\alpha\beta}  {\partial_\mu \eta^\dagger_c}\bar D^*_\nu  \overlrarrow{\partial}_\alpha D^*_\beta +\up{H.c.}, \\
=&+2g_{\psi  D D} \psi^{\dagger\mu} \bar D  {\partial}_\mu D    \\
& - \i 2 g_{\psi DD^*} \frac{1}{M_\psi} \epsilon^{\mu\nu\alpha\beta}  {\partial_\mu \psi^\dagger_\nu}( \bar D  {\partial}_\alpha D^*_\beta   +\bar D^*_\alpha  { \partial}_\beta D   ) \\
& +g_{\psi D^* D^*} \psi^{\dagger\mu} (\bar D^{*\nu}  {\partial}_\nu D^*_\mu  -  {\partial}_\nu \bar D^*_\mu   D^{*\nu} +  2{\partial}_\mu \bar D^{*\nu}  D^{*}_\nu ) \\
&+g_{D D^*\eta_c}  \eta^\dagger_c ({\partial_\mu} \bar D D^{*\mu} - \bar D^{*\mu} {\partial_\mu} D) \\
& + \i 2 g_{D^* D^*\eta_c} \frac{1}{M_{\eta_c}}\epsilon^{\mu\nu\alpha\beta}  {\partial_\mu \eta^\dagger_c}\bar D^*_\nu   {\partial}_\alpha D^*_\beta +\up{H.c.},
\end{aligned}
\end{equation}
}
\begin{equation} \label{E-L-All}
\begin{aligned}
\mathcal{L}_{2} =&+g_{\psi  D D} \psi^{\dagger\mu} \bar D  {\partial}_\mu D    \\
& - \i g_{\psi DD^*} \frac{1}{M_\psi} \epsilon^{\mu\nu\alpha\beta}  {\partial_\mu \psi^\dagger_\nu}( \bar D  {\partial}_\alpha D^*_\beta   +\bar D^*_\alpha  { \partial}_\beta D   ) \\
& +g_{\psi D^* D^*} \psi^{\dagger\mu} (\bar D^{*\nu}  {\partial}_\nu D^*_\mu  -  {\partial}_\nu \bar D^*_\mu   D^{*\nu} +  2{\partial}_\mu \bar D^{*\nu}  D^{*}_\nu ) \\
&+g_{D D^*\eta_c}  \eta^\dagger_c ({\partial_\mu} \bar D D^{*\mu} - \bar D^{*\mu} {\partial_\mu} D) \\
& + \i g_{D^* D^*\eta_c} \frac{1}{M_{\eta_c}}\epsilon^{\mu\nu\alpha\beta}  {\partial_\mu \eta^\dagger_c}\bar D^*_\nu   {\partial}_\alpha D^*_\beta +\up{H.c.},
\end{aligned}
\end{equation}
where we have divide a meson mass in the second and the last Lagrangians to keep all the coupling constants dimensionless. The symbol $\epsilon^{\mu\nu\alpha\beta}$ denotes the totally antisymmetric  Levi-Civita tensor with $\epsilon^{\mu\nu\alpha\beta}=-\epsilon_{\mu\nu\alpha\beta}$ and convention $\epsilon^{0123}=1$. All these coupling constants are related to a single coupling $g_2$, which is determined to be $g_2 = {\sqrt{M_\psi}}/({2M_D f_\psi})$ with $f_\psi$ denoting the $J/\psi$ decay constant\,\cite{Colangelo2004}.  
Then all other coupling constants can also be expressed in terms of the $g_{\psi D D}$ as
\begin{equation} 
\begin{aligned}
\textstyle g_{\psi D D} &= \textstyle  \frac{2M_\psi}{f_\psi},\\
\textstyle g_{\psi D D^*}    &=\textstyle\left(\frac{M_{D^*}}{M_D} \right)^{1/2} \textstyle g_{\psi D D}, \\
\textstyle g_{\psi D^* D^*} &= \textstyle \frac12 \left( \frac{M_{D^*}}{M_D} \right)g_{\psi D D}, \\
\textstyle g_{D D^* \eta_c}&=\textstyle \frac12 \left( \frac{M_{\eta_c} M_{D^*}}{M_\psi M_D} \right)^{1/2}g_{\psi D D}, \\
\textstyle g_{D^* D^* \eta_c}&=\textstyle\left( \frac{M_{\eta_c}}{M_\psi } \right)^{1/2} \frac{M_{D^*}}{M_D}g_{\psi D D}.
\end{aligned}
\end{equation}
In next section, we will also discuss the effects of these coupling constants on the final decay widths.

\subsection{Amplitude for $P_\psi^N(4312)^+ \to J/\psi p$}

$P_\psi^N(4312)^+$ as the $\bar D \Sigma_c$ molecular state can decay to $J/\psi p$ by  exchanging either a  $D$ or a $D^*$ virtual meson, and the total amplitude is the sum of the two.

\begin{figure}[htpb]
\centering
\includegraphics[width = 0.45\textwidth, angle=0]{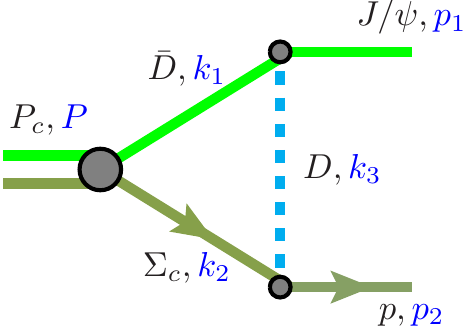}
\includegraphics[width = 0.45\textwidth, angle=0]{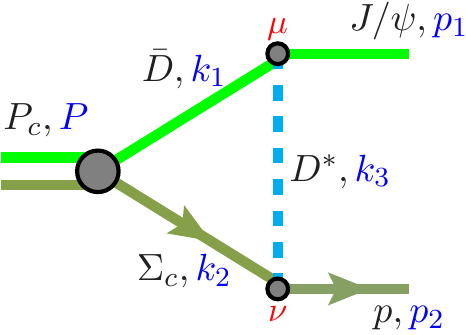}
\caption{Strong decay of $P_\psi^N(4312)^+$ to the $J/\psi p$ by exchanging a virtual mediator $D$\,(left panel) and $D^*$\,(right panel). $P,~k_1,~k_2, ~P_1,~P_2$ denote the momenta of $P_\psi^N$, constituent meson,  constituent baryon, the final $J/\psi$, and the final $p$ respectively.}\label{Fig-Pc-psi+p-D}
\end{figure}

\subsubsection{Amplitude with $D$ exchange}

The left panel of \autoref{Fig-Pc-psi+p-D} shows the Feynman diagram of $P_\psi^N(4312)^+ \to J/\psi p$ by exchanging the $D$ meson.
Besides the pentaquark vertex, we also need two other effective Lagrangian to obtain the decay width.
From above results, the effective Lagrangian describing the $DDJ/\psi$ interaction read
\begin{gather}
\mathcal L_{\psi DD} = g_{\psi D D} (\psi^{\mu})^\dagger  \bar D \partial_\mu D ,
\end{gather}
where $D=(D^0, D^+,D_s^+)^\up{T}$ represents the charmed meson fields in flavor triplet, and $\bar D$ represents fields of the corresponding anti-charmed mesons. Whereas the effective Lagrangian for $N\!D\Sigma_c$ interaction behaves as\,\cite{Garzon2015,XiaoCJ2019}
\begin{gather}
\mathcal{L}_{N\!D\Sigma_c} =( -\i) g_{\!N\!D\Sigma_c} \bar N \gamma_5 \Sigma_c \bar D^\dagger + \up{H.c.}, 
\end{gather}
where $N$ stands for nucleon field doublet; $\Sigma_c=\bm{\sigma}\cdot\bm{ \Sigma_c}$ with $\bm \sigma$ denoting the Pauli matrix and $\bm{\Sigma_c}$ denoting the $\Sigma_c$ baryon isospin triplet. 

The invariant amplitude for $P_\psi^N(4312)^+\to J/\psi p$ by exchanging a $D$ can then be expressed by the Bethe-Salpeter vertex as
\begin{gather}
\mathcal{A}_1 = \int \frac{\d^4 k}{(2\pi)^4} \bar u_2 (-\i g_{N\!D\Sigma_c})\gamma^5 \left[S(k_2) \Gamma(P,k,r) D(k_1) \right] (g_{\psi D D}) D(k_3) (\i)e^{*\alpha}_{1} k_{1\alpha},
\end{gather}
where $u_2$ is short for $u_{(r_2)}(P_2)$ with $r_2$ representing the proton spin state; $e_{1}$ is short for $e_{(r_1)}(P_1)$ representing the polarization vector of the final $J/\psi$ with $P_1$ denoting the $J/\psi$ momentum and $r_1=0,\pm 1$ representing the 3 possible polarization states. The polarization vector $e_{1}$ fulfills the Lorentz condition
\begin{gather}
e_{1}^\alpha  P_{1\alpha} =0.
\end{gather}
The momentum of the exchanged virtual charmed meson is denoted as $k_3=(k_1-P_1)$.
We will use $M_1$ and $M_2$ to denote the masses of the final $J/\psi$ and proton respectively.

Also notice $k_1=(\alpha_1 P+k)$ and $k_3$ are involved in the four-dimensional integration over $k$. To simplify this amplitude, first we strip off the triangle amplitude involved the integral over $k$ as
\begin{gather} \label{E-AmpT}
{T}_{1\alpha} u(P,r)= \gamma_5 \int \frac{\d^4 k}{(2\pi)^4} \left[S(k_2) \Gamma(P,k,r) D(k_1) \right]  D(k_3)  k_{1\alpha},
\end{gather} 
where the Lorentz condition of the vector meson is utilized, and we also strip off the spinor $u(P,r)$ for later convenience. The decay amplitude $\mathcal{A}_1$ can then be simplified as
\begin{gather}
\mathcal{A}_1 =    (g_{N\!D\Sigma_c} g_{\psi DD}) e^{*\alpha}_{1}   [\bar u_{2}   {T}_{1\alpha}  u(P,r)].
\end{gather}
Then we perform the contour integral over $k_P$ on Eq.\,(\ref{E-AmpT}), and obtain
\begin{align} \label{E-A-alpha-k}
 {T}^\alpha_{1} u(P,r)&=\gamma_5 \int \frac{\d k^3_\perp}{(2\pi)^3} \frac{1}{2w_3}( a^\alpha_1 \varphi^+  + a^\alpha_2 \varphi^- ), 
\end{align}
where we used the expression of the positive(negative) energy wave functions $\varphi^\pm=\Lambda^\pm \gamma^0 \varphi$; the two coefficients $a_1$ and $a_2$ behaves as
\begin{gather}
a^\alpha_1 = c_1 x_1^\alpha + c_2 x_2^\alpha + c_3 x_3^\alpha, \\
a^\alpha_2 = c_4 x_4^\alpha + c_5 x_5^\alpha + c_6 x_6^\alpha, 
\end{gather} 
where $x_i = k_1(k_P=k_{Pi})$ with $(i=1,\cdots 6)$, and $k_{Pi}$s are defined as
\begin{gather}
k_{P1}=\zeta_1^+,~k_{P2}=\zeta_2^+,~ k_{P3}=\zeta^+_3,~k_{P4}=\zeta_1^-,~ k_{P5}=\zeta_2^-,~ k_{P6}=\zeta_3^-,
\end{gather}
where the abbreviations $\zeta_1^\pm\equiv -(\alpha_1 M \mp w_1)$, $\zeta_2^\pm \equiv( \alpha_2 M\mp w_2)$, and
 $\zeta_3^\pm \equiv (E_1-\alpha_1M \pm w_3) $ are used. 
The coefficients $c_i$s $(i=1,\cdots 6)$ are defined as 
\begin{equation} \label{E-cn}
\begin{aligned}
c_{1(4)} &= \frac{1}{w_1+w_3\mp E_1},\\
c_{2(5)} &= \frac{(-1)}{w_2+w_3\mp E_2}, \\
c_{3(6)} &= \frac{(w_1+w_2\mp M )}{(w_1+w_3 \pm E_1)(w_2+w_3 \mp E_2)}.
\end{aligned}
\end{equation}
Now the amplitude $\mathcal{A}_{1T;\alpha}$ has been expressed by the three-dimensional Salpeter wave function $\varphi(k_\perp)$,  and can be further simplified as the two form factors
\begin{gather}
{T}_{1\alpha} = (s_{1P} \gamma_\alpha  + s_{2P} \hat P_\alpha ),
\end{gather}
Inserting the obtained $P_\psi^N(4312)^+$ wave function, namely, \eref{E-wave1-1-2N}, into \eref{E-A-alpha-k}  and then calculating the three-dimensional integral numerically, we can obtain the amplitude $T_{1\alpha}$ in terms of $s_{1P}$ and $s_{1P}$.
In the appendix \ref{App} we collect the specific expressions of the two form factors in terms of the Salpeter wave functions $f_1$ and $f_2$.
The amplitude $\mathcal A_1$ then behaves as
\begin{gather}
\mathcal{A}_1 = (g_{N\!D\Sigma_c} g_{\psi DD}) e^{*\alpha}_{1}   \bar u_{2}  (s_{1P} \gamma_\alpha  + s_{2P} \hat P_\alpha) u(P,r).
\end{gather}

\subsubsection{Amplitude with $D^*$ exchange}

For $P_\psi^N(4312)^+$ decaying by exchanging $D^*$ in the lowest level, the relevant Feynman diagram is displayed in the right panel of \autoref{Fig-Pc-psi+p-D}, and the two involved interaction vertexes are $DD^*J/\psi$ and $\Sigma_cD^*p$. The $DD^*J/\psi$ interaction is represented by the following effective Lagrangian
\begin{gather}
\mathcal{L}_{\psi DD^*} =(- \i) g_{\psi D\!D^*} \epsilon^{\mu\nu\alpha\beta} \frac{1}{M_\psi}\partial_\mu \psi^\dagger_\nu  \partial_\alpha D^*_\beta   \bar D  ,
\end{gather}
Notice here the coupling constant $g_{\psi DD^*}$ is defined to have the same dimension with the pseudoscalar coupling constant $g_{\psi DD}$.  The effective Lagrangian of $ND^*\Sigma_c $ reads\,\cite{LiuW2001,Garzon2015,YangF2107}
\begin{gather} \label{E-L-BBDStar}
\mathcal{L}_{N\!D^*\Sigma_c} = \i g_{N\!D^*\Sigma_c } \bar N \gamma^\alpha  \Sigma_c D^{*\dagger}_\alpha  + \up{H.c.}.
\end{gather}
All the coupling constants in these effective Lagrangian will be specified in next section.

The invariant amplitude for $P_\psi^N(4312)^+\to J/\psi p$ by exchanging a $D^*$ can be expressed by the Bethe-Salpeter vertex as
\begin{align}
\mathcal{A}_2 &=(g_{N\!D^*\Sigma_c} g_{\psi D D^*}) \bar u_{2}  \gamma^\nu  
\int \frac{\d^4 k}{(2\pi)^4} \left[S(k_2) \Gamma(P,k,r) D(k_1) \right] D_{\mu\nu}(k_3) \frac{-\i^3}{M_1} \epsilon^{P_1 \alpha \beta \mu} e^*_{1\alpha}  k_{1\beta}, 
\end{align}
where the propagator of the exchanged $D^*$ meson behaves as
\begin{gather}
D_{\mu\nu}(k_3) = \i \frac{-g_{\mu\nu} + {k_{3\mu} k_{3\nu}}/{m_3^2}}{k_3^2 - m_3^2 + \i \epsilon},
\end{gather}
Here the propagator mass $m_3$ is $M_{D^*}$. 
Notice the contraction with Levi-Civita tensor forces the momentum part in numerator of $D_{\mu\nu}(k_3)$ to be zero. The amplitude can be further simplified as
\begin{align}
\mathcal{A}_2 =(g_{\Sigma_cN\!D^*} g_{\psi D D^*}) e^{*\alpha}_{1}  \bar u_{2}  \gamma^\mu (-\i)\frac{1}{M_1}\epsilon_{P_1\alpha \beta \mu}  {T}_{2}^\beta u(P,r), 
\end{align}
where we have stripped off the amplitude involved the integral over $k$ as before
\begin{align}\label{E-AmT2}
{T}_{2}^\beta u(P,r)= \int \frac{\d^4 k}{(2\pi)^4} \left[S(k_2) \Gamma(P,k,r) D(k_1) \right] D(k_3)  k_{1}^\beta .
\end{align}
In order to express the amplitude by the three-dimensional Salpeter wave function, we perform the contour integration over $k_P$ on \eref{E-AmT2} as usual and obtain 
\begin{align} \label{E-A2T}
{T}_{2}^\beta u(P,r) 
&= \int \frac{\d^3 k_\perp}{(2\pi)^3} \frac{1}{2w_3} \left( a_1^{\beta}  \varphi^+  +  a_2^{\beta} \varphi^-\right).
\end{align}
Combining ${T}_{2}^\beta u(P,r)$ with $\gamma^\mu \frac{(-\i)}{M_1}\epsilon_{P_1\alpha \beta \mu}$, and using the following identity of the Levi-Civita symbol,
\begin{align}
\i \gamma_\mu \epsilon^{\mu \alpha\beta\nu} = \gamma^5(\gamma^\alpha \gamma^\beta \gamma^\nu - \gamma^\alpha g^{\beta\nu} +\gamma^\beta g^{\alpha\nu} - \gamma^\nu g^{\alpha\beta}),
\end{align}
we can express the decay amplitude $\mathcal{A}_2$ for  $D^*$ exchange by the following form factors
\begin{align}
\mathcal{A}_2 
&= (g_{\psi DD^*}g_{\Sigma_cN\!D^*}) e_1^{*\alpha}\bar u_2  (s_{1V} \gamma_\alpha + s_{2V} \hat P_\alpha) u(P,r).
\end{align}
Namely, the amplitude $\mathcal{A}_2$ can be expressed by the same form as $\mathcal{A}_1$, which is just what it should be. 
In above equations, the specific expressions of $s_{1V}$ and $s_{2V}$ can be obtained by inserting the Salpeter wave functions into \eref{E-A2T} and performing the integral numerically. The specific expressions are presented in the appendix\,\ref{App}.

\subsection{Amplitude for $P_\psi^N(4312)^+ \to \eta_c p$}
Since $\eta_c$ with $J^P=0^-$, the decay to $\eta_c p$ can only happen by exchanging a $D^*$  while the mode of exchanging a $D$ is forbidden. From \eref{E-L-All}, the effective Lagrangian responsible for $DD^*\eta_c$ interaction reads
\begin{align}
\mathcal{L}_{DD^*\eta_c}  
&=  g_{D\!D^*\eta_c}  \eta^\dagger_c \partial_\mu \bar D D^{*\mu}.
\end{align}
The effective Lagrangian describing $ND^*\Sigma_c$ interaction has been presented in \eref{E-L-BBDStar}. 
The corresponding Feynman diagram is similar with that for the decay to $J/\psi p$ with $D^*$ exchange. The decay amplitude for $P_\psi^N(4312)^+\to \eta_c p$ behaves as
\begin{align}
  \mathcal{A}_3 & = \bar u_2  (\i g_{N\!D^*\Sigma_c} )\gamma^\alpha \int \frac{\d^4 k}{(2\pi)^4} [S(k_2) \Gamma(k,r) D(k_1)] D_{\alpha\beta}(k_3)g_{ D^*D\eta_c}(-\i P_1^\beta ).
\end{align}
As usual, it is convenient to strip off the part involved the integral over $k$ as
\begin{align*}
{T}_3 u(P,r)
& = \int \frac{\d^4 k}{(2\pi)^4}(\gamma^\alpha  P_1^\beta)[S(k_2) \Gamma(k,r) D(k_1)] D_{\alpha\beta}(k_3),
\end{align*}
Performing the contour integral over $k_P$, we can express ${T}_3$ by the three-dimensional Salpeter wave function
\begin{align}
{T}_{3} u(P,r)
&=  \int \frac{\d^3 k_\bot}{(2\pi)^3}\frac{1}{2w_3} (\gamma^\alpha  P_1^\beta) \sum_{i=1}^3 \left[  c_i d_{\alpha\beta}(y_i) \varphi^+   +c_{i+3} d_{\alpha\beta}(y_{i+3}) \varphi^-   \right],
\end{align}
where the positive\,(negative) energy wave function is related to the Salpeter wave function by $\varphi^\pm = \Lambda^\pm \gamma_0 \varphi$; and we define $d_{\alpha\beta}$ and $y_i$ as
\begin{equation}\label{E-y}
\begin{gathered}
d_{\alpha\beta}(y_i) = -g_{\alpha\beta} + \frac{y_{i\alpha} y_{i\beta}}{m_3^2},\\
y_i =k_3(k_P=k_{Pi}) =x_{iP} \hat P + k_\bot - P_1.
\end{gathered}
\end{equation}
Notice that contribution of the momentum part in $d_{\alpha\beta}$ will be suppressed when the exchanged particle is heavy. 
Inserting the Salpeter wave function \eref{E-wave1-1-2N} of $P_\psi^N(4312)^+$, we obtain $T_3$ expressed by one form factor,
\begin{gather}
T_3 = s_{3V} \gamma_5.
\end{gather}
Finally,  we obtain the amplitude for decay to $\eta_c p$ by form factor $s_{3V}$ with a simple form
\begin{gather} \label{E-Am-etac}
 \mathcal{A}_3  = \left(g_{\!N\!D^*\Sigma_c} g_{\eta_c D\!D^*} \right) \bar u_2  \left(s_{3V}\gamma_5 \right) u(P,r).
\end{gather}
The expression of $s_{3V}$ is also listed in appendix\,\ref{App} as the integral over Salpeter wave functions.

\subsection{$P_\psi^N(4312)^+\to \bar D^{*0}\Lambda_c^+$}
The strong decay of $P_\psi^N(4312)\to \bar D^{*0}\Lambda_c^+$ is similar with the decay to $J/\psi p$, just the vector meson $J/\psi$ replaced by $\bar D^{*0}$, the proton replaced by the $\Lambda_c^+$ baryon, and the propagator $D^{(*)}$ replaced by the $\pi(\rho)$ respectively.
The effective Lagrangian describing the interaction of $\bar D^* \bar D \phi$ and $\Lambda_c \Sigma_c \phi$ are\,\cite{Oh2001,DingGJ0809,ShenCW2016,LinYH2017} 
\begin{gather}
\mathcal{L}_{\bar D\bar D^*\phi} = g_{\bar D \bar D^*\phi} ( \bar D^{*\mu})^\dagger  \partial_\mu \phi \bar D ,\\
\mathcal{L}_{ \Sigma_c\Lambda_c \phi} =( -\i) g_{\Lambda_c\Sigma_c\phi} \bar \Lambda_c  \gamma_5 \Sigma_c \phi,
\end{gather}
where  $\phi$ is the $3\times3$ traceless hermitian matrix consisting of eight pseudo-scalar meson fields, 
\begin{gather}
\phi=
\begin{bmatrix}
\frac{\pi^0}{\sqrt2}+ \frac{\eta}{\sqrt6} & \pi^+  & K^+ \\
\pi^-  & -\frac{\pi^0}{\sqrt2}+ \frac{\eta}{\sqrt6} & K^0 \\
K^-  & \bar K^0 & -\frac{2}{\sqrt6} \eta
\end{bmatrix}.
\end{gather}
The coupling constants can be obtained under the heavy quark spin-flavor symmetry. Here we use the coupling constant $g_{\bar D\bar D^*\phi}=\frac{2g}{f_\pi}\sqrt{M_DM_{D^*}}$ with the $\pi$ decay constant $f_\pi=0.132\,\si{GeV}$ and the coupling constant $g=0.59$\,\cite{DingGJ0809}, and the coupling constant $g_{\Lambda_c\Sigma_c\phi} = 19.3$\,\cite{LinYH2017}.

Combining above effective Lagrangian and the BS vertex, we can express the decay amplitude of $P_\psi^N(4312)^+\to \bar D^{*0}\Lambda_c^+$ by exchanging a $\pi$ as
\begin{gather}
\mathcal{A}_4 =  \int \frac{\d^4 k}{(2\pi)^4} (-\i g_{\Lambda_c \Sigma_c \phi}) \bar u_2 \gamma_5  [S(k_2) \Gamma(k,r) D(k_1)] D(k_3)  (g_{D^*D\phi})(\i k_{1\alpha}) (e_1^{\alpha})^*.
\end{gather}
By taking a similar calculation procedure with that in decay to $J/\psi p$, we can further express this amplitude by two form factors,
\begin{gather}
\mathcal{A}_4 = (g_{\Lambda_c \Sigma_c \phi} g_{D^*D\phi}) (e_1^{\alpha})^* \bar u_2 \left (s_{4P} \gamma_\alpha + s_{5P}\hat P_\alpha \right) u(P,r),
\end{gather} 
where the form factor $s_{4P}$ and $s_{5P}$ has exactly the same expressions with $s_{1P}$ and $s_{2P}$, respectively,  just the masses of $m_3$, $M_1$, and $M_2$ changed from $M_D$, $M_{J/\psi}$, and $M_p$ to $M_\pi$, $M_{\bar D^{*0}}$, and $M_{\Lambda_c}$, respectively.

The decay to $\bar D^{*0}\Lambda_c^+$ by exchanging a $\rho$ is similar with the decay to $J/\psi p$ by exchanging a $D^*$, and the relevant effective Lagrangian are
\begin{gather}
\mathcal{L}_{\bar D\bar D^* V} = -g_{\bar D \bar D^*V} \frac{\i }{M_{\bar D^*}} \epsilon^{\mu\nu\alpha\beta} \partial_\mu (\bar D^*_\nu)^\dagger \partial_\alpha {V_\beta} \bar D, \\
\mathcal{L}_{\Sigma_c\Lambda_c V} =\i  g_{\Sigma_c\Lambda_c V} \bar \Lambda_c \gamma_\alpha \Sigma_c {V}^\alpha,
\end{gather}
where the coupling constants $g_{\bar D \bar D^*V}=4\rho_\up{T1}=9.3\,\si{GeV}^{-1}$, and $g_{\Sigma_c\Lambda_c V}=0.56$\,\cite{LinYH2017}.
Then we can express the corresponding amplitude as
\begin{align}
\mathcal{A}_5 &=(g_{\Lambda_c\Sigma_cV} g_{\bar D\bar D^*V}) \bar u_{2}  \gamma^\alpha  
\int \frac{\d^4 k}{(2\pi)^4} \left[S(k_2) \Gamma(P,k,r) D(k_1) \right] D_{\alpha\beta} (k_3) \frac{(-\i^3)}{M_1} \epsilon^{P_1 \mu \nu \beta} e^*_{1\mu} (\i k_{3\nu}). 
\end{align}
Here the propagator mass $m_3$ is $M_{\rho}$. Again by taking a similar calculation procedure as in subsection 3.1.2, the invariant amplitude is finally expressed by the form factors
\begin{align}
\mathcal{A}_5
&= (g_{D^*D\rho}g_{\Lambda_c \Sigma_c\rho}) e_1^{*\alpha}\bar u_2  (s_{4V} \gamma_\alpha + s_{5V} \hat P_\alpha) u(P,r),
\end{align}
where the form factor $s_{4(5)V} = s_{1(2)V}(m_3=M_\rho,M_1=M_{\bar D^{*0}},M_{2}=M_{\Lambda_c})$.

Finally,  we can express the total amplitude for decay to $\bar D^{*0}\Lambda_c^+$ as
\begin{gather}
\mathcal{A}[P_\psi^N(4312)^+{\bar D^*\Lambda_c}] = \mathcal{A}_4 +\mathcal{A}_5
\end{gather}

\subsection{$P_\psi^N(4312)^+\to \bar D^{0}\Lambda_c^+$ by exchanging a $\rho$}
The decay of  $P_\psi^N(4312)^+ \to \bar D^{0}\Lambda_c^+$ is quite similar with the decay to $\eta_c p$. The involved interaction Lagrangian is
\begin{gather}
\mathcal{L}_{\bar D\bar DV} = g_{\bar D\bar DV} \partial_\mu (\bar D^\dagger)  {V}^\mu  \bar D, 
\end{gather}
where the coupling constant $g_{\bar D\bar D V}=2\rho_\up{V1}=7.51$.
The corresponding decay amplitude behaves as
\begin{align}
  \mathcal{A}_6 & = g_{\Lambda_c\Sigma_cV} \bar u_2 \gamma^\alpha \int \frac{\d^4 k}{(2\pi)^4} [S(k_2) \Gamma(k,r) D(k_1)] D_{\alpha\beta}(k_3)g_{\bar D\bar DV} (\i P_1^\beta ).
\end{align}
As usual, it is convenient to strip off the part involved the integral over $k$ as
\begin{align*}
T_6 u(P,r)
& = \int \frac{\d^4 k}{(2\pi)^4}(\gamma^\alpha  P_1^\beta)[S(k_2) \Gamma(k,r) D(k_1)] D_{\alpha\beta}(k_3),
\end{align*}
which can be further simplified by finishing the integral involved the Salpeter vertex as 
\begin{gather}
T_6 = s_{6V}\gamma_5,
\end{gather}
where the only form factor $s_{6V}$ has the same expression with $s_{3V}$, just the taking the parameter values $m_3=M_\rho,M_1=M_{\bar D^{0}},M_{2}=M_{\Lambda_c}$ in the form factor expressed by the Salptere wave functions.
Finally,  the decay amplitude of  $P_\psi^N(4312)^+ \to \bar D^{0}\Lambda_c^+$ can be expressed as
\begin{gather}  \label{E-Am6}
 \mathcal{A}_6  =  \i ( g_{\Lambda_c\Sigma_c V} g_{\bar D\bar DV} ) \bar u_2  \left(s_{6V}\gamma_5\right) u(P,r).
\end{gather}

\subsection{Partial decay widths}
 
Combing the two amplitudes from $D$ and $D^*$ mediators together, we obtain the full invariant amplitude for $P_\psi^N(4312)^+\to J/\psi p$ decay by two form factors,
\begin{gather} \label{E-A}
\mathcal{A} [P_\psi^N(4312)^+\to J/\psi p]
=  \mathcal{A}_1+\mathcal{A}_2 =  e^{*\alpha}_{1}  \bar u_{2} \left(   s_1 \gamma_\alpha + s_2 \hat P_\alpha \right) u(P,r).
\end{gather}
where $s_1$ and $s_2$ are related to the coupling constants and are expressed as 
\begin{equation}\label{E-s1-s2}
\begin{aligned}
s_1&=g_{\psi \!D\!D}g_{\!N\!D\Sigma_c}s_{1P} + g_{\psi \!D\!D^*}g_{\!N\!D^*\Sigma_c} s_{1V},\\
s_2&=g_{\psi \!D\!D}g_{\!N\!D\Sigma_c}s_{2P} + g_{\psi \!D\!D^*}g_{\!N\!D^*\Sigma_c} s_{2V}.
\end{aligned}
\end{equation}
Similarly, the total invariant amplitude for $P_\psi^N(4312)^+\to \bar D^{*0}\Lambda_c^+$ can also be obtained as
\begin{gather} \label{E-A-3}
\mathcal{A} [P_\psi^N(4312)^+\to \bar D^{*0}\Lambda_c^+]
=  \mathcal{A}_4+\mathcal{A}_5=  e^{*\alpha}_{1}  \bar u_{2} \left(   s_4 \gamma_\alpha + s_5 \hat P_\alpha \right) u(P,r).
\end{gather}
where $s_4$ and $s_5$ are related to the coupling constants and are expressed as 
\begin{equation}\label{E-s4-s5}
\begin{aligned}
s_4&=g_{ \bar D \phi \bar D^*}g_{\Sigma_c \phi \Lambda_c}s_{4P} + g_{\bar D V \bar D^*}g_{\Sigma_c V  \Lambda_c} s_{4V},\\
s_5&=g_{ \bar D \phi \bar D^*}g_{\Sigma_c  \phi  \Lambda_c}s_{5P} + g_{\bar D V \bar D^*}g_{\Sigma_c V \Lambda_c} s_{5V}.
\end{aligned}
\end{equation}

For the decays with final vector meson $J/\psi$ or $\bar D^{*0}$, squaring the amplitude and summing all the polarization states, we obtain 
\begin{gather}
\sum_{r_1,r_2,r} |\mathcal{A}|^2 =  \left( -g^{\alpha\beta}+\frac{P_1^\alpha P_1^\beta}{M_1^2} \right)\up{Tr}  \left(\slashed P_{\!2}+M_2\right)  T_{\alpha}[J/\psi(\bar D^{*0})] \left( \slashed P+M \right) \bar{T}_{\beta},
\end{gather} 
where 
\begin{gather}
T_{\alpha}[J/\psi(\bar D^{*0})]  = \left(   s_{1(4)} \gamma_\alpha + s_{2(5)}\hat P_\alpha \right),
\end{gather}
and $\bar{T}_{\beta} = \gamma^0 T_{\beta}^\dagger \gamma_0$ is defined as the usual conjugation variable;  we also used the relationship of the summation over the vector polarization states $r_1$,
\begin{gather}
 \sum_{r_1} e_{(r_1)}^\alpha e_{(r_1)}^{*\beta} = -g^{\alpha\beta}+\frac{P_1^\alpha P_1^\beta}{M_1^2};
\end{gather} 
and the summation over the polarization states of the spinors
\begin{gather}
 \sum_{r_2} u_{(r_2)}(P_2) \bar u_{(r_2)}(P_2) =  (\slash P_2 + M_2),\\
  \sum_r u(P,r) \bar u(P,r) =  (\slashed P + M).
\end{gather}

For $P_\psi^N(4312)^+\to \eta_c p(\bar D^0 \Lambda_c^+)$ decay, the squared amplitude behaves as
\begin{gather}
\sum_{r_2,r} |\mathcal{A}_{3(6)}|^2 = \up{Tr}  \left(\slashed P_{\!2}+M_2\right)  \mathcal{A}_{3(6)} \left( \slashed P+M \right) \bar{\mathcal{A}}_{3(6)} = 4 M (E_2-M_2) s_{3(6)}^2,
\end{gather} 
where $s_3=g_{{N\!D^*\Sigma_c}} g_{\eta_c D^*D} s_{3V}$ and $s_6=g_{\Lambda_c\Sigma_c V} g_{\bar D\bar DV} s_{6V}$.
The squared amplitude is proportional to the kinetic energy of the final baryon. 

Finally, the partial decay width of $P_\psi^N(4312)^+$ to $J/\psi(\eta_c) p$ or $\bar D^{(*)0}\Lambda_c^+$ is expressed as
\begin{gather}
\Gamma[P_\psi^N(4312)^+ \!\to \! M_{P(V)}B] = \frac{|\bm{P}_1|}{8\pi M^2}  C_\up{I}\frac{1}{2} \sum_{r,r_1,r_2} |\mathcal{A}|^2 ,
\end{gather} 
where $C_\up{I}$ denotes the isospin factor, and the three momentum of the final meson, $J/\psi(\eta_c)$ or $\bar D^{(*)0}$, is given by
\begin{gather}
|\bm{P}_1| =\frac{1}{2M}\left[\Big(M^2-(M_1+M_2)^2\Big) \Big(M^2-(M_1-M_2)^2\Big)\right]^{\frac12}.
\end{gather}

\section{Numerical results and discussions}\label{Sec-4}
\subsection{Numerical parameters}
Before giving the decay widths, we first summarize the effective interaction coefficients we used in above effective Lagrangian. The interaction coefficients between the heavy hadron and the light bosons are obtained under the heavy quark symmetry, which read\,\cite{YangZC2012,ChenR2015,ChenR2019,HeJ2019,XuH2020}: $\rho_\up{V1}=\frac{\beta g_V}{\sqrt2}=3.75$, $\rho_\up{T1}=\frac{\lambda g_V}{\sqrt2}=2.34\,\si{GeV}^{-1}$, and $\sigma_1=0.76$; $\rho_\up{V2}=\frac{\beta_Sg_V}{\sqrt2}=7.26$, $\rho_\up{T2}=\frac{\lambda_Sg_V}{\sqrt2}=13.81\,\si{GeV}^{-1}$, and $\sigma_2=6.2$.
In the heavy quark limit, the coupling constants between the heavy hadrons read\,\cite{Colangelo2004}
$g_{D\!D\psi}=\frac{2M_\psi}{f_\psi}$ and $g_{D\!D^*\psi}= (\frac{ M_{D^*}}{M_D})^{1/2}g_{D\!D\psi}$ with the $J/\psi$ decay constant $f_\psi=0.416\,\si{GeV}$ estimated from the dilepton decay width\,\cite{PDG2020}; the $DD^*\eta_c$ coupling constant reads $g_{D\!D^*\eta_c}=\frac12(\frac{M_{\eta_c}M_{D^*}}{M_\psi M_D})^{1/2}g_{D\!D\psi}$. Combined with the total amplitude \eref{E-A}, it can be found that the partial decay width is proportional to $\frac{1}{f_\psi^2}$. The coupling constants related to the baryons used are $g_{\!N\!D\Sigma_c}=2.69$ and $g_{\!N\!D^*\Sigma_c}=3.0$\,\cite{Garzon2015,XiaoCJ2019}. These values are the standard parameters used in this work, and we will also vary the standard parameters to explore their influence on the wave functions and the final decay widths.

The hadron masses used are $M_{P_\psi^N(4312)^+}=4.312\,\si{GeV}$, $M_\psi=3.097\,\si{GeV}$, $M_{\eta_c}=2.983\,\si{GeV}$, $M_p=0.938\,\si{GeV}$, $M_{\bar D^{*0}}=2.007\,\si{GeV}$, $M_{\Lambda_c^+}=2.286\,\si{GeV}$\,\cite{PDG2022}.

\subsection{Numerical results and theoretical uncertainties}
\begin{figure}[h!]
\vspace{0.5em}
\centering
\subfigure[]{\includegraphics[width=0.48\textwidth]{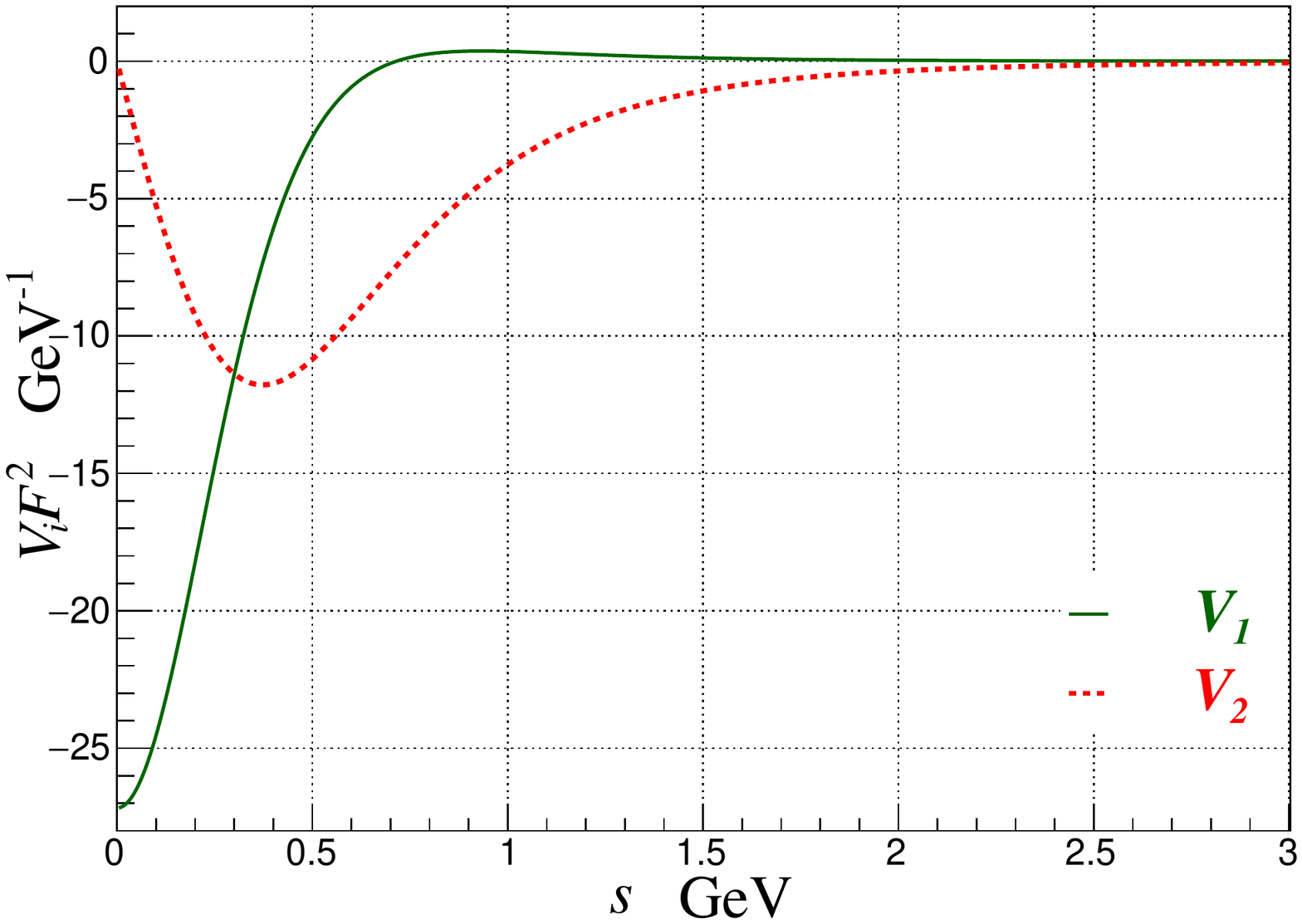} \label{Fig-V2}}
\subfigure[]{\includegraphics[width=0.48\textwidth]{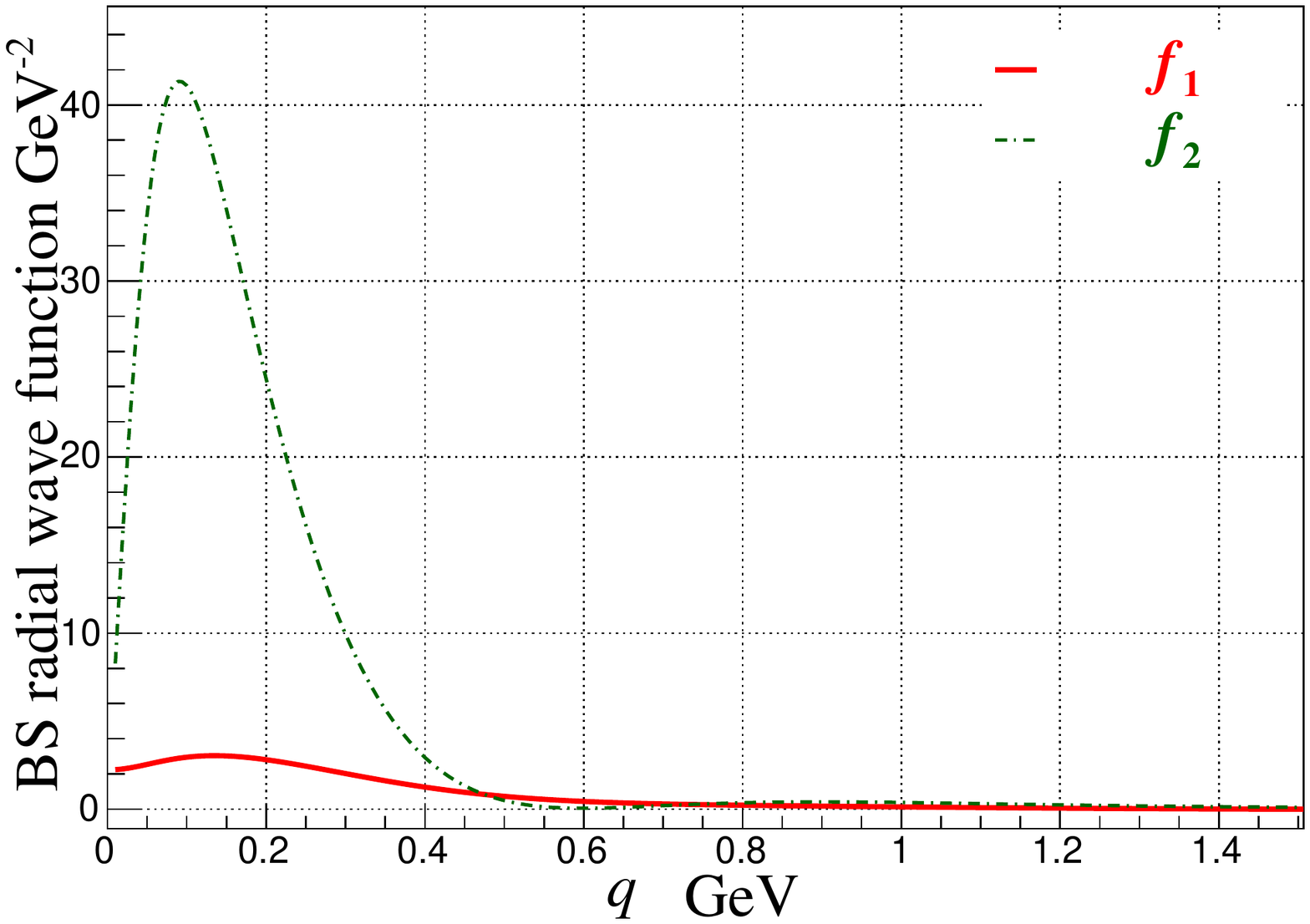} \label{Fig-wave-0}}
\subfigure[]{\includegraphics[width=0.48\textwidth]{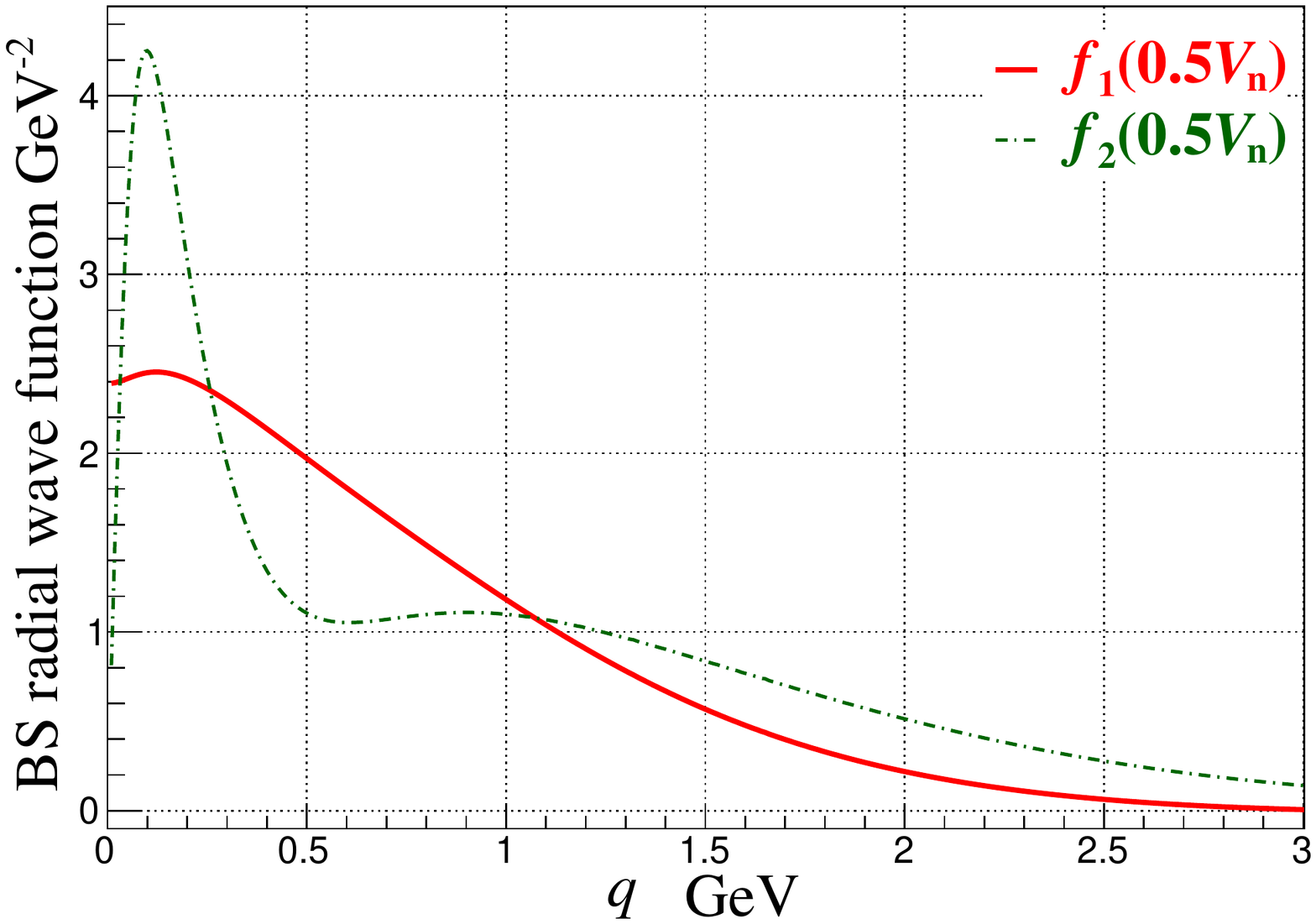} \label{Fig-wave-E1}}
\subfigure[]{\includegraphics[width=0.48\textwidth]{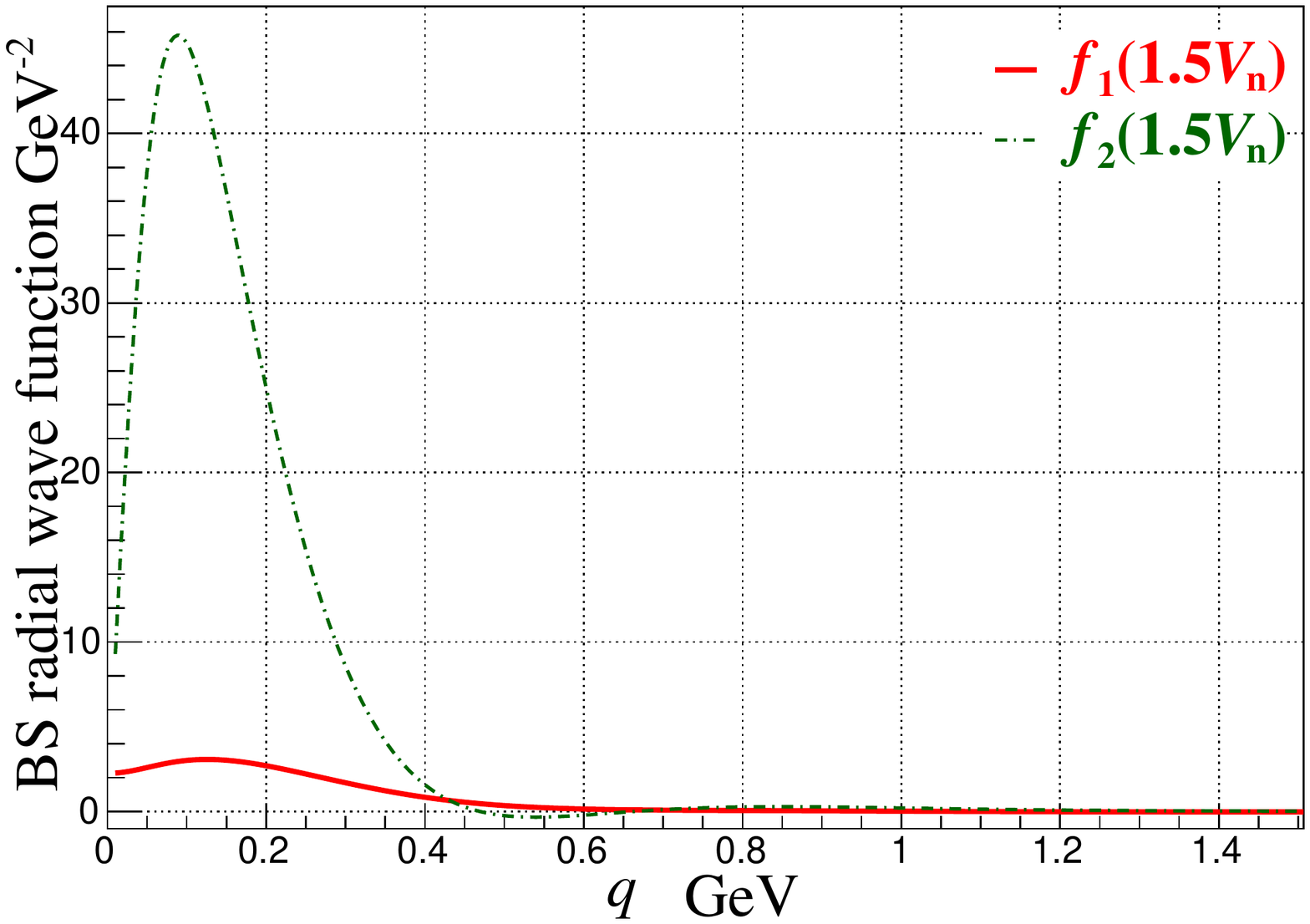} \label{Fig-wave-E2}}
\caption{The figure (a) shows the isospin-$\frac12$ potentials $V_n F^2~(n=1,2)$. Figure (b) is the Bethe-Salpeter radial wave function $f_1$ and $f_2$ for  $P_\psi^N(4312)^+$ as the $\bar D\Sigma_c$ molecular state based on the one-boson exchange; while (c) and (d) are the radial wave functions when the interaction potential $V_n$ in \eref{E-KVB-Vn} is reduced and increased by 50\% based, respectively, where the corresponding regulator values are $m_\Lambda=1.288\,\si{GeV}$ and 0.73\,\si{GeV} respectively.}\label{Fig-Vn+wave}
\vspace{0.5em}
\end{figure}
The only free parameter in this work is the regulator $m_\Lambda$ in the form factor $F(\bm{s}^2)$ in \eref{E-form-factor}. All the other parameters have been determined by the previous experimental data. By solving the relevant BS eigenvalue equation, we find proper cutoff values of $m_\Lambda$ can produce bound state of $\bar D\Sigma_c$ based on the one-boson exchange kernel in  isospin-$\frac12$. Then by fitting the bound state mass $M$ to the experimental measurement $M_{P_\psi^N(4312)^+}=4.312\,\si{GeV}$,  we fix $m_\Lambda$ in \eref{E-form-factor} to be $0.87\,\si{GeV}$. Then the obtained $V_1$ and $V_2$ in the interaction kernel are graphically shown in \autoref{Fig-V2}.

In  \autoref{Fig-wave-0} we show the obtained BS wave functions $f_1$ and $f_2$ for $P_\psi^N(4312)^+$.  On the other hand, the obtained radial wave functions depend on the obtained potential $V_{1(2)}$, which is directly related to the coupling constants $\sigma_{1(2)}$, $\rho_\up{V1(2)}$ and $\rho_\up{T2}$. To reflect the influence of these parameters on the wave functions and decay widths, we vary the numerical values of $V_{1(2)}$ under standard parameters by $\mp 50\%$. Under these variations, the obtained regulator values are then $m_\Lambda=1.288\,\si{GeV}$ and 0.73\,\si{GeV} respectively, and the corresponding wave functions obtained are displayed in \autoref{Fig-wave-E1} and \autoref{Fig-wave-E2}. As $V_{1(2)}$ decreases, the fitted regulator parameter $m_\Lambda$ increases, and also the role of wave function $f_1$ becomes more important.

Our  results of the mass spectra for $I(J^P)=1/2(1/2)^-$ $\bar D\Sigma_c$ molecule indicate that there only exists one bound state, namely, $P_\psi^N(4312)^+$ as the ground state of $\bar D\Sigma_c$ molecule. Our results do not support any radially excited states. This conclusion is robust even under the $\pm50\%$ change of the interaction kernel $V_{1(2)}$. 

The obtained numerical values of the form factors for decays to $J/\psi p$ and $\bar D^{*0}\Lambda^+_c$ channels in \eref{E-A} are
\begin{gather*}
 s_{1P}=1.1\times10^{-3},~s_{2P}=1.9\times10^{-3},~s_{1V}=-4.1\times10^{-3},~s_{2V}=7.6\times10^{-3}; \\
 s_{4P}=7.5\times 10^{-3},~s_{5P} = 6.0 \times10^{-3},~ s_{4V} = -8.7 \times10^{-3},~s_{5V} = 1.2\times10^{-2}.
\end{gather*}
For decays to $\eta_c p$ and $\bar D^0 \Lambda_c^+$, the obtained form factors $s_{3V}$ in \eref{E-Am-etac}  and $s_{6V}$ in \eref{E-Am6}  are 
\begin{gather*}
s_{3V}=5.1\times10^{-3},~~s_{6V} = 3.6\times 10^{-2}.
\end{gather*}

Inserting above form factors into the decay width expressions, we obtain the partial decay widths as $\Gamma[P_\psi^N(4312)^+ \!\to \!J/\psi p]  = 0.11\,\si{MeV} $ and $\Gamma[P_\psi^N(4312)^+ \!\to \!\eta_c p] = 0.085\,\si{MeV}$.
The obtained partial width for decay to $\!J/\psi p$ is $\sim\!1\%$ of the total width  $\Gamma_\up{Tot}=9.8\pm2.7^{+3.7}_{-4.5}$ MeV\,\cite{LHCb2019-Pc} reported by the LHCb collaboration.  The $\!J/\psi p$ channel is also the only observed decay mode of $P_\psi^N(4312)^+$ currently. While the decay fraction of $P_\psi^N(4312)^+$ to $\eta_c p$ is about $50\%$ smaller than the $J/\psi p$ channel. There is still no evident signal in recent experimental search of $P_\psi^N(4312)^+$ in $\eta_c p$ channel\,\cite{LHCb2020-Lambdab}. Notice the obtained results are totally predictive and there are no any free adjustable parameters since the regulator $m_\Lambda$ has been fixed by the mass of $P_\psi^N(4312)^+$. 
\begin{table}[ht]
\caption{Comparison of  partial decay widths of  $P_\psi^N(4312)^+$ to $J/\psi(\eta_c) p$ and $\bar D^{(*)0} \Lambda_c^+$ with other works in units of MeV, where our theoretical uncertainties are induced by varying the relevant coupling constants by $\pm 5\%$ in the effective Lagrangian.
}\label{Tab-Comparison}
\vspace{0.2em}\centering
\begin{tabular}{ c|ccccccccccc }
\toprule[1.5pt]
Channel 		    & This	    						   &\cite{WangGJ2020} & \cite{XuYJ2020} 	  		&\cite{LinYH2019}     &\cite{DongYB2020}	 & \cite{XiaoCJ2019}	 &\cite{WangZG2020}	 \\
 \midrule[1.2pt]
$J/\psi p$ 	  	    &  $0.17^{-0.04}_{+0.04}$    		   &$0.32\pm0.08$	 &$1.67^{+0.92}_{-0.56}$ 	&$10^{-3}\sim0.1$ 	&$0.033$	& $(3\sim8)$	 		 &$9.3_{-9.3}^{+19.5}$ 	 \\
$ \eta_c p$ 	  	    &  $0.085^{-0.016}_{+0.018}$    &$0.98\pm0.25$	 &$5.54^{+0.75}_{-0.50}$ 	&$10^{-2}\sim0.4$ 	&$0.066$	& $-$	 		   		 &$0.26_{+0.55}^{-0.24}$  \\
$\bar D^{*0}\Lambda_c^+$ &$8.8^{-1.6}_{+1.9}$ & -  & - & 10.7 &  6.16 &-&-\\
$\bar D^{0}\Lambda_c^+$  &$0.026^{-0.05}_{+0.06}$& - & - & 0.3    &  0      &-&-\\
\bottomrule[1.5pt]
\end{tabular}
\end{table}

Our result of $P_\psi^N(4312)^+\!\to\! \bar D^{*0}\Lambda_c^+$ is $8.8\,\si{MeV}$, which can amount to $\sim90\%$ of the total width. This prediction is consistent with the calculations in Refs.\,\cite{LinYH2019,DongYB2020}. The partial decay width $\Gamma[P_\psi^N(4312)^+\!\to\! \bar D^{0}\Lambda_c^+]$ is about $0.026\,\si{MeV}$, which is comparable with decay  to $\eta_c p$ but negligible with $\bar D^{*0}\Lambda_c^+$. The huge difference between $\bar D^{*0}$ and $\bar D^0$ channel mainly stems from the difference of the involved coupling constants. Our results support $\bar D^{*0}\Lambda_c^+$ to be the dominant channels for $P_\psi^N(4312)^+$, which would be a most promising decay channel to be detected in experiments. The sum of these four decay widths are $\sim9.1\,\si{MeV}$, which amounts to $\sim93\%$ of the total width $\Gamma=9.8\,\si{MeV}$ reported by the LHCb.

A comparison of our results with other works is listed in \autoref{Tab-Comparison}. Our obtained partial decay widths are roughly consistent with those in Refs.\,\cite{WangGJ2020,LinYH2019,DongYB2020}.
Notice the theoretical results for decay widths of $P_\psi^N(4312)\!\to\! J/\psi(\eta_c) p$ are quite different from each other for the complication of this problem. Also besides the result in Ref.\,\cite{WangZG2020}, the partial decay width to $\eta_c p$ is great than $J/\psi p$ in Refs.\,\cite{WangGJ2020,XuYJ2020,LinYH2019,DongYB2020}, which is roughly consist with a simple analysis in the heavy quark limit in Ref.\,\cite{Voloshin2019}. However, our result of the partial decay width to $\eta_c p$ is about the half of that to $J/\psi p$. The reason for this difference are analyzed as follows. First, the phase space of $J/\psi p$ channel is almost the same with that of $\eta_c p$ channel. While after summing over all polarization states, we obtain the squared amplitudes as
\begin{gather}
\sum_{r,r_1,r_2} |\mathcal{A}(J/\psi p)|^2=  17.5s_1^2 + 8.2s_1s_2 + 1.6s_2^2, \\
\sum_{r,r_1,r_2} |\mathcal{A}(\eta_c p)|^2=5.0s_3^2,
\end{gather}
where the form factor $s_1$ part dominates in  $J/\psi p$ channel, which corresponds to the $J/\psi p$ channel amplitude obtained in Ref.\,\cite{Voloshin2019} in the nonrelativistic format. It is easy to see
the coefficient in front of $s_1^2$ in $J/\psi p$ channel is $\sim3$ times of that in $\eta_c p$ channel, which steps from the vector polarization states of $J/\psi$. On the other hand, the coupling constants $g_{\psi DD^{(*)}}$ is about 2 times of $g_{\eta_c DD^*}$; and the obtained form factor $s_{3V}$ for $\eta_c p$ decay channel is roughly near with the dominant item $s_{1V}$ for $J/\psi p$, which finally makes our partial decay width for $\eta_c p$ channel is smaller than that for $J/\psi p$ channel. Notice in the $\bar D^{(*)}\Lambda_c$ decay channels, the obtained form factor $s_{6V}$ in pseudoscalar mode is about $3$ times of that in the vector mode, which is consistent with the nonrelativistic estimation in the heavy quark limit in Ref.\,\cite{Voloshin2019}. The main difference between $J/\psi(\eta_c)p$ and $\bar D^{(*)}\Lambda_c$ channels stems from the much heavier exchanged particle $D^*$ in the former one, which would suppress the contribution of the momentum part in the numerator of the propagator $D_{\alpha\beta}(k_3)$ and then make $s_{3V}\sim s_{1V}$. However, more researches are still needed.

Since the obtained partial decay widths are also directly dependent on the coupling constants $g_{\psi D\!D}$, $g_{\psi D\!D^*}$, $g_{N\!D\Sigma_c}$, $g_{N\!D^*\Sigma_c}$, and $g_{D\!D^*\eta_c}$ in the relevant effective Lagrangian. To see the sensitivity of the our partial decay widths on these parameters, we calculate the theoretical uncertainties by varying the every coupling constant by $\pm 5\%$, and then searching the parameter space to find the maximum deviation. The obtained theoretical errors are also listed in above \autoref{Tab-Comparison}, where the relative uncertainties induced from the coupling constants amount to about $\sim20\%$ for the four channels we calculated here. Another theoretical uncertainties come from the interaction strength in the BS kernel, which can be collected into the kernel potential $V_{1(2)}$. When the interaction kernel $V_{1(2)}$ varies by $\pm50\%$ based on the standard parameters, the obtained decay widths are 4.3\,\si{MeV} and $0.04\,\si{MeV}$ for $J/\psi p$ channel, respectively; while for $\bar D^{*0}\Lambda_c^+$ channel, the results are $1.1\,\si{MeV}$ and $9.1\,\si{MeV}$, respectively; while for $\bar D^{0}\Lambda_c^+$ channel, the results are $2.5\,\si{MeV}$ and $0.005\,\si{MeV}$, respectively.

{Also it is worthy to notice that the obtained partial decay widths for $J/\psi(\eta_c)p$ channels are both proportional to $\frac{1}{f_\psi^2}$, while the values of $f_\psi$ used in Ref.\,\cite{XiaoCJ2019} and Ref.\,\cite{XuYJ2020} are $0.426$ and $0.481\,\si{GeV}$ respectively, under which our result for $J/\psi(\eta_c) p$ channel would be $5\%$ and $25\%$ smaller respectively. }

\subsection{Summary}

We give a brief summary. In this work, firstly, based on the effective Lagrangian in the the heavy quark limit, we calculate the one-boson-exchange interaction kernel of $\bar D\Sigma_c$ in the isospin-$\frac12$ state. Then by using the Bethe-Salpeter equation, we obtain the mass spectrum and wave functions of the experimental $P_\psi^N(4312)^+$ as the $\bar D\Sigma_c$ molecular state  with $J^P=(\frac{1}{2})^-$. Then combining the effective Lagrangian and the obtained BS wave function, we calculate the partial decay width to be $0.17\,\si{MeV}$, $0.085\,\si{MeV}$, $8.8\,\si{MeV}$, and $0.026\,\si{MeV}$ for $P_\psi^N(4312)^+\!\to\! J/\psi$,  $\eta_c p$, $\bar D^{*0}\Lambda_c^+$ and $\bar D^0\Lambda_c^+$, respectively. The obtained numerical results indicate that the fraction of $\bar D^{*0}\Lambda_c^+$ channel can amount to $\sim90\%$ of $P_\psi^N(4312)^+$, which makes $\bar D^{*0}\Lambda_c^+$ to be a much more promising decay channel to be discovered in experiments. This result can also serve as an important test for the molecular interpretation of $P_\psi^N(4312)^+$. Our results are roughly consistent with some other calculations and also the LHCb experimental measurements. However, more theoretical analysis and experimental measurements are necessary to determine the properties of the pentaquark state $P_\psi^N(4312)^+$.
The interpretation of  $P_\psi^N(4312)^+$  as the $\bar D\Sigma_c$ molecular state with $J^P=(\frac{1}{2})^-$  and isospin $I=\frac12$ is favored by this work.

\appendix

\section{Expressions of the decay form factors} \label{App}
For completeness, we list the specific expressions of the relevant form factors here, which are all represented by the integral over the radial Salpeter wave functions $f_1$ and $f_2$. Parts of following expressions are calculated with the help of the \texttt{FeynCalc} package\,\cite{Mertig1990,Shtabovenko2016,Shtabovenko2020}.
The four form factors for decay $P_\psi^N(4312)^+\!\to\! J/\psi p$ in \eref{E-s1-s2} are
\begin{equation}
\begin{aligned}
s_{1P} &=-\int \frac{\d^3 k_\perp}{(2\pi)^3} \frac{1}{2w_2} c_{22}  k\left[ k v_0 f_1 + \left(w_2 u_0+m_2
   v_0\right) f_2 \right], \\
s_{2P} &=\int \frac{\d^3 k_\perp}{(2\pi)^3} \frac{1}{2w_2P_1^2} \left( Y_1f_1 + Y_2 f_2 \right),   \\
s_{1V} &=-\int \frac{\d^3 k_\perp}{(2\pi)^3} \frac{1}{2w_2P_1^2} \left( Z_1f_1 + Z_2 f_2 \right),   \\
s_{2V} &=-\int \frac{\d^3 k_\perp}{(2\pi)^3} \frac{1}{2w_2P_1^2} \left( Z_3f_1 + Z_4 f_2 \right),   \\
\end{aligned}
\end{equation}
where $Y_1$, $Y_2$, and $Z_1\sim Z_4$ read 
\begin{equation}
\begin{aligned}
Y_1 =& P_1^2 w_2 u_1+c_1 E_1 k P_1 v_1+c_1 k M_2 P_1 v_1-c_1
   k M P_1 v_1-m_2 P_1^2 v_1-c_1 E_1 k P_1 w_2 u_0\\
   &-c_{21}
   E_1^2 k^2 v_0-c_{21} E_1 k^2 M_2 v_0+c_{21}
   E_1 k^2 M v_0+c_1 E_1 k m_2 P_1 v_0+c_{22} k^2
   P_1^2 v_0  , 
\end{aligned}
\end{equation}
\begin{equation}
\begin{aligned}    
Y_2 =& c_1 E_1 P_1 w_2 u_1-c_1 M P_1 w_2 u_1+c_1 M_2 P_1 w_2
   u_1-c_1 E_1 m_2 P_1 v_1-2 c_1 {E_2} m_2 P_1 v_1 ~~\\
   &+c_1
   m_2 M_2 P_1 v_1+c_1 m_2 M P_1 v_1+k P_1^2 v_1-c_{21}
   E_1^2 k w_2 u_0+c_{21} E_1 k M w_2 u_0 \\
   &-c_{21}
   E_1 k M_2 w_2 u_0+c_{22} k P_1^2 w_2 u_0+c_{21}
   E_1^2 k m_2 v_0+2 c_{21} E_1 {E_2} k m_2
   v_0 \\
   &-c_1 E_1 k^2 P_1 v_0-c_{21} E_1 k m_2 M_2
   v_0-c_{21} E_1 k m_2 M v_0+c_{22} k m_2 P_1^2 v_0,
\end{aligned}
\end{equation}
\begin{equation}
\begin{aligned}   
Z_1=& -E_1 P_1^2 w_2 u_1+M P_1^2 w_2 u_1+M_2 P_1^2 w_2 u_1-c_1
   E_1^2 k P_1 v_1+c_1 k M_1^2 P_1 v_1+E_1 m_2 P_1^2
   v_1 \\
   &-m_2 M P_1^2 v_1-m_2 M_2 P_1^2 v_1+c_1 E_1^2 k P_1
   w_2 u_0-c_1 E_1 k M P_1 w_2 u_0-c_1 E_1 k M_2 P_1
   w_2 u_0 \\
   &+c_{21} E_1^3 k^2 v_0-c_1 E_1^2 k m_2 P_1
   v_0-c_{22} E_1 k^2 P_1^2 v_0-c_{21} E_1 k^2 M_1^2
   v_0 \\
   &+c_1 E_1 k m_2 M P_1 v_0+c_1 E_1 k m_2 M_2 P_1
   v_0-c_{22} k^2 M P_1^2 v_0-c_{22} k^2 M_2 P_1^2 v_0  ,
\end{aligned}
\end{equation}
\begin{equation}
\begin{aligned}      
Z_2 =& c_1 M_1^2 P_1 w_2 u_1+c_1
   E_1^2 m_2 P_1 v_1+2 c_1 E_1 {E_2} m_2 P_1
   v_1-2 c_1 E_1 m_2 M P_1 v_1 \\
   &+c_1 m_2 M_1^2 P_1
   v_1+k M P_1^2 v_1+k M_2 P_1^2 v_1+c_{21}
   E_1^3 k w_2 u_0+c_1 E_1^2 k^2 P_1 v_0\\
   &-c_{21} E_1 k M_1^2 w_2 u_0-c_{22}
   E_1 k P_1^2 w_2 u_0-c_{22} k M P_1^2 w_2 u_0-c_{22} k
   M_2 P_1^2 w_2 u_0\\
   &+2
   c_{21} E_1^2 k m_2 M v_0-c_1 E_1 k^2 M P_1 v_0-c_1
   E_1 k^2 M_2 P_1 v_0+c_{22} E_1 k m_2 P_1^2
   v_0 \\
   &-c_{21} E_1 k m_2 M_1^2 v_0+2 c_{22} {E_2} k m_2
   P_1^2 v_0-3 c_{22} k m_2 M P_1^2 v_0-c_{22} k m_2 M_2 P_1^2
   v_0  \\
   & -2 c_{21}E_1^2 {E_2} k m_2 v_0-c_{21} E_1^3km_2 v_0 -E_1 k P_1^2 v_1 -c_1 E_1^2 P_1 w_2 u_1,
\end{aligned}
\end{equation}
\begin{equation}
\begin{aligned}      
Z_3=& -M P_1^2 w_2 u_1-M_2 P_1^2 w_2 u_1+c_1 E_1 k M P_1 v_1-c_1
   E_1 k M_2 P_1 v_1-c_1 k M_1^2 P_1 v_1 \\
   &+m_2 M_2 P_1^2 v_1+c_1 E_1 k M P_1 w_2 u_0+c_1
   E_1 k M_2 P_1 w_2 u_0-c_{21} E_1^2 k^2 M
   v_0 \\
   &+c_{21} E_1^2 k^2 M_2 v_0+c_{21} E_1 k^2 M_1^2
   v_0-c_1 E_1 k m_2 M P_1 v_0-c_1 E_1 k m_2 M_2 P_1
   v_0\\
   &+m_2 M P_1^2
   v_1-c_{22} k^2 M_2 P_1^2 v_0 +3 c_{22} k^2 M P_1^2 v_0  ,
\end{aligned}
\end{equation}
\begin{equation}
\begin{aligned}      
Z_4=& c_1 E_1 M P_1 w_2 u_1-c_1 E_1 M_2 P_1 w_2 u_1-c_1
   M_1^2 P_1 w_2 u_1+c_1 E_1 m_2 M P_1 v_1\\
   &-c_1 m_2 M_1^2 P_1 v_1-k M P_1^2 v_1-k M_2
   P_1^2 v_1-c_{21} E_1^2 k M w_2 u_0+c_{21} E_1^2 k
   M_2 w_2 u_0 \\
   &+c_{21} E_1 k M_1^2 w_2 u_0+3 c_{22} k M
   P_1^2 w_2 u_0-c_{22} k M_2 P_1^2 w_2 u_0-c_{21} E_1^2
   km_2 M v_0\\
   &+c_{21} E_1^2 k m_2 M_2 v_0 +c_1 E_1
   k^2 M P_1 v_0+c_1 E_1 k^2 M_2 P_1 v_0+c_{21} E_1 k
   m_2 M_1^2 v_0\\
   &+3 c_{22} k m_2 M P_1^2 v_0-c_{22} k m_2 M_2
   P_1^2 v_0-c_1 E_1
   m_2 M_2 P_1 v_1 .
\end{aligned}
\end{equation}
In above expressions, $P_1=|\bm{P}_1|$, and 
\begin{align}
c = \cos\theta,~~c_{21}= \frac{1}{2} (3\cos^2\theta -1),~~c_{22}= \frac{1}{2} (\cos^2\theta-1),
\end{align}
where $\theta$ denotes the angle between $\bm k$ and $\bm{P}_1$. We also define $u_{n}$ and $v_{n}\,(n=0,1,2)$  for later convenience
\begin{equation}
\begin{aligned}
u_{n} &= (c_1x_{1P}^n+c_2 x_{2P}^n+c_3x_{3P}^n) + (c_{4}x_{4P}^n+c_5 x_{5P}^n+c_6x_{6P}^n),\\
v_{n} &= (c_1x_{1P}^n+c_2 x_{2P}^n+c_3x_{3P}^n) - (c_{4}x_{4P}^n+c_5 x_{5P}^n+c_6x_{6P}^n).
\end{aligned}
\end{equation}
The expressions of $c_i$ are listed in \eref{E-cn}.

The form factor $s_{3V}$ in \eref{E-Am-etac} for $P_\psi^N(4312)^+\!\to \eta_c p$ decay behaves
\begin{equation}
\begin{aligned}
s_{3V}=\int \frac{\d^3 k_\perp}{(2\pi)^3} \frac{1}{4 m_3^2 w_2 w_3P_1^2} \left[ (P_1^2X_1 + kc_1X_3) f_1  + (k P_1^2X_2 + c_1X_4)f_2 \right],
\end{aligned}
\end{equation}
where $X_1\sim X_4$ read
\begin{equation}
\begin{aligned}
X_1=&-c k M P_1 w_2 u_0-c k M_2 P_1 w_2 u_0+m_3^2 M w_2 u_0+m_3^2 M_2
   w_2 u_0-M M_1^2 w_2 u_0 \\
   &-M_1^2 M_2 w_2 u_0+c k m_3 P_1 w_2
   u_{{1}}+E_1 m_3 M w_2 u_{{1}}+E_1 m_3
   M_2 w_2 u_{{1}}+m_3 M_1^2 w_2 u_{{1}} \\
   &-E_1
   m_3^2 w_2 u_{{2}}+c k^3 P_1 v_0+k^2 M_1^2 v_0+c k m_2 M
   P_1 v_0+c k m_2 M_2 P_1 v_0 \\
   &-c k m_2 m_3 P_1 v_{{1}}-m_2
   m_3^2 M v_0+m_2 M M_1^2 v_0-m_2 m_3^2 M_2 v_0+m_2 M_1^2 M_2
   v_0 \\
   &-E_1 k^2 m_3 v_{{1}}-E_1 m_2 m_3 M
   v_{{1}}-E_1 m_2 m_3 M_2 v_{{1}}+E_1
   m_2 m_3^2 v_{{2}}-m_2 m_3 M_1^2 v_{{1}}, 
\end{aligned}
\end{equation}   
\begin{equation}
\begin{aligned}
X_2 =&c k P_1 w_2 u_0+M_1^2 w_2 u_0-E_1 m_3 w_2 u_{{1}}+c k
   m_2 P_1 v_0+c k m_3 P_1 v_{{1}}+m_3 M_1^2 v_{{1}} \\
   &-c k M P_1 v_0-c k M_2
   P_1 v_0+m_3^2 M v_0+m_3^2 M_2 v_0+m_2 M_1^2 v_0-M M_1^2
   v_0 \\
   &-M_1^2 M_2 v_0+E_1 m_3 M v_{{1}}+E_1 m_3
   M_2 v_{{1}}-E_1 m_2 m_3 v_{{1}}-E_1
   m_3^2 v_{{2}} ,
\end{aligned}
\end{equation}   
\begin{equation}
\begin{aligned}   
X_3 =& -c E_1 k P_1 w_2 u_0-E_1 M_1^2 w_2 u_0+c k M P_1 w_2
   u_0+c k M_2 P_1 w_2 u_0+M M_1^2 w_2 u_0 \\
   &+M_2 M_1^2 w_2
   u_0+E_1^2 m_3 w_2 u_{{1}}-E_1 m_3 M w_2
   u_{{1}}-E_1 m_3 M_2 w_2 u_{{1}}+c E_1
   k m_2 P_1 v_0 \\
   &-c E_1 k m_3 P_1 v_{{1}}-c E_1
   k M P_1 v_0-c E_1 k M_2 P_1 v_0+E_1 m_2 M_1^2
   v_0+E_1 m_3^2 M v_0 \\
   &+E_1 m_3^2 M_2 v_0-E_1 M
   M_1^2 v_0-E_1 M_2 M_1^2 v_0+c k m_3 M P_1
   v_{{1}}+c k m_3 M_2 P_1 v_{{1}} \\
   &-c k m_2 M P_1
   v_0-c k m_2 M_2 P_1 v_0+c k M_1^2 P_1 v_0-m_3^2 M_1^2 v_0-m_2
   M M_1^2 v_0 \\
   &-m_2 M_2 M_1^2 v_0+M_1^4 v_0+E_1^2 m_3 M
   v_{{1}}+E_1^2 m_3 M_2 v_{{1}}-E_1^2
   m_2 m_3 v_{{1}}+E_1^2 m_3^2 v_{{2}}\\
   &-2E_1 m_3 M_1^2 v_{{1}}+E_1 m_2 m_3 M
   v_{{1}}+E_1 m_2 m_3 M_2 v_{{1}}-E_1
   m_3^2 M v_{{2}}-E_1 m_3^2 M_2 v_{{2}} \\
   &+m_3 M
   M_1^2 v_{{1}}+m_3 M_2 M_1^2 v_{{1}}     ,
\end{aligned}
\end{equation}   
\begin{equation}
\begin{aligned}   
X_4 =& v_0 m_2 M_1^4+u_0 w_2 M_1^4-v_0 m_2 m_3^2 M_1^2+E_1 k^2 v_0
   M_1^2+2 E_2 k^2 v_0 M_1^2 +m_2 m_3 M_2
   v_{{1}} M_1^2\\
   &-k^2 M v_0 M_1^2-E_1 M
   v_0 m_2 M_1^2+k^2 v_0 M_2 M_1^2-E_1 v_0 m_2 M_2 M_1^2+c
   k v_0 m_2 P_1 M_1^2 \\
   &-u_0 m_3^2 w_2 M_1^2-E_1 M u_0 w_2
   M_1^2-E_1 u_0 M_2 w_2 M_1^2+c k u_0 P_1 w_2 M_1^2+2
   E_2 m_3 w_2 u_{{1}} M_1^2 \\
   &-M m_3 w_2 u_{{1}}
   M_1^2+m_3 M_2 w_2 u_{{1}} M_1^2-2 E_1 m_2 m_3
   v_{{1}} M_1^2+M m_2 m_3 v_{{1}} M_1^2 \\
   &+E_1 M v_0 m_2 m_3^2+E_1 v_0
   m_2 m_3^2 M_2+c E_1 k^3 v_0 P_1+2 c E_2 k^3 v_0
   P_1-c k^3 M v_0 P_1 \\
   &-c E_1 k M v_0 m_2 P_1+c k^3 v_0 M_2
   P_1-c E_1 k v_0 m_2 M_2 P_1+E_1 M u_0 m_3^2
   w_2+E_1 u_0 m_3^2 M_2 w_2 \\
   &-c E_1 k M u_0 P_1 w_2-c
   E_1 k u_0 M_2 P_1 w_2+E_1^2 M m_3 w_2
   u_{{1}}+E_1^2 m_3 M_2 w_2 u_{{1}}+c
   E_1 k m_3 P_1 w_2 u_{{1}}\\
   &+2 c E_2 k m_3 P_1
   w_2 u_{{1}}-c k M m_3 P_1 w_2 u_{{1}}+c k m_3 M_2
   P_1 w_2 u_{{1}}-E_1^2 m_3^2 w_2 u_{{2}}-2
   E_1 E_2 m_3^2 w_2 u_{{2}} \\
   &+E_1 M m_3^2
   w_2 u_{{2}}-E_1 m_3^2 M_2 w_2
   u_{{2}}-E_1^2 k^2 m_3 v_{{1}}-2 E_1
   E_2 k^2 m_3 v_{{1}}+E_1 k^2 M m_3
   v_{{1}} \\
   &+E_1^2 M m_2 m_3 v_{{1}}-E_1
   k^2 m_3 M_2 v_{{1}}+E_1^2 m_2 m_3 M_2
   v_{{1}}-c E_1 k m_2 m_3 P_1 v_{{1}}+c k M
   m_2 m_3 P_1 v_{{1}} \\
   &+c k m_2 m_3 M_2 P_1
   v_{{1}}+E_1^2 m_2 m_3^2 v_{{2}}-E_1 M
   m_2 m_3^2 v_{{2}}-E_1 m_2 m_3^2 M_2 v_{{2}}.
\end{aligned}
\end{equation}

\acknowledgments
The author Q. Li thanks Prof.\,Fen-Kun Guo of ITP-CAS, and Dr.\,Xu-Chang Zheng of Chongqing Univ., and Dr. Hao Xu of Northwest Normal Univ. for helpful suggestions and discussions. This work is supported by the National Natural Science Foundation of China\,(NSFC) under Grant Nos.\,12005169, 12075301, 11821505, 12047503, 11805024, 11865001, and 12075073. It is also supported by the National Key R\&D Program of China\,(2022YFA1604803), the Natural Science Basic Research Program of Shaanxi\,(Program No.\,2021JQ-074), and the Fundamental Research Funds for the Central Universities.



\end{document}